\documentclass[preprint2]{aastex}

\def\hinv{$h^{-1}$}
\def\gtapr {\lower .1ex\hbox{\rlap{\raise .6ex\hbox{\hskip .3ex
        {\ifmmode{\scriptscriptstyle >}\else
                {$\scriptscriptstyle >$}\fi}}}
        \kern -.4ex{\ifmmode{\scriptscriptstyle \sim}\else
                {$\scriptscriptstyle\sim$}\fi}}}
\def\ltapr {\lower .1ex\hbox{\rlap{\raise .6ex\hbox{\hskip .3ex
        {\ifmmode{\scriptscriptstyle <}\else    
                {$\scriptscriptstyle <$}\fi}}}
        \kern -.4ex{\ifmmode{\scriptscriptstyle \sim}\else
                {$\scriptscriptstyle\sim$}\fi}}}
\begin{document}
\title{The CNOC2 Field Galaxy Redshift Survey I: The Survey and the Catalog
for the Patch CNOC$\,$0223+00}
\author{
H.~K.~C.~Yee\altaffilmark{1,2}, S.L.~Morris\altaffilmark{3,2}, 
H.~Lin, \altaffilmark{1,2,4,5},
R.G.~Carlberg\altaffilmark{1,2},
P.B. Hall\altaffilmark{1,2},
Marcin Sawicki\altaffilmark{1,2,6},
D.R.~Patton\altaffilmark{1,2,7},
Gregory D.~Wirth\altaffilmark{2,7,8},
E.~Ellingson\altaffilmark{9,2},
and C.W.~Shepherd\altaffilmark{1}
}

\altaffiltext{1} {Department of Astronomy, University of Toronto, Toronto, 
Ontario M5S 3H8, Canada. Email: hyee,
carlberg, hall, patton, and shepherd @astro.utoronto.ca}
\altaffiltext{2}{Visiting Astronomer, Canada-France-Hawaii Telescope,
which is operated by the National Research Council of Canada, Le Centre 
National de Recherche Scientifique, and the University of Hawaii.}  
\altaffiltext{3}{Dominion Astrophysical Observatory, Herzberg Institute
of Astrophysics, National Research Council, 5071 W. Sannich Rd, 
Victoria, BC, V8X 4M6, Canada.  Email: Simon.Morris@hia.nrc.ca}
\altaffiltext{4}{Present address: Steward Observatory, University of
Arizona, Tucson, AZ  85721. Email: hlin@as.arizona.edu}
\altaffiltext{5}{Hubble Fellow}
\altaffiltext{6}{Present address: CalTech, Mail Code 320--47,
 Pasadena, CA 91125. Email:
sawicki@pirx.caltech.edu}
\altaffiltext{7}{Department of Physics and Astronomy, University of
Victoria, Victoria, BC, V8W 3P6}
\altaffiltext{8}{Present address: Keck Observatory, Waimea, HI 96743.
Email: wirth@keck.hawaii.edu}
\altaffiltext{9}{CASA, University of Colorado, Boulder, CO 80309.
Email: e.elling@casa.colorado.edu}

\received{}
\accepted{}

\begin{abstract}
The Canadian Network for Observational Cosmology (CNOC2) 
Field Galaxy Redshift Survey is a spectroscopic/photometric
survey of faint galaxies over 1.5 square degrees of sky with a nominal
spectroscopic limit of $R_c\sim21.5$ mag.
The primary goals of the survey are to investigate the evolution
of galaxy clustering and galaxy populations over the
redshift range of $\sim 0.1$ to $\sim0.6$.
The survey area contains four widely separated patches on the sky with
a total sample of over $6000$ redshifts, representing a sampling
rate of about 45\%.
In addition,
5-color photometry (in $I_c$, $R_c$, $V$, $B$, and $U$)
for a complete sample of approximately 40,000 galaxies 
to $R_c\sim23.0$ mag is also available.
We describe the survey and observational strategies, multi-object
spectroscopy mask design procedure, and data reduction
 techniques for creating the spectroscopic-photometric catalogs.
We also discuss the derivations of various  statistical weights, including
corrections for the effects of limited spectral bandwidth, for the 
redshift sample which allow it to be used as a complete sample.
As the initial release of the survey data,
we present the full data set and some statistics 
for the Patch CNOC$\,$0223+00.

\end{abstract}
 \keywords{
galaxies: redshifts ---
galaxies: photometry ---
galaxies: general ---
surveys ---
techniques: photometric ---
techniques: spectroscopic
}
 
\section{Introduction}

Fundamental to our understanding of the Universe is the formation and
evolution of structures, from galaxies to clusters of galaxies 
to large-scale structures such as sheets, filaments, and voids.
Theoretical advances, often in the form of increasingly larger and  more
sophisticated N-body simulations, have laid much of the groundwork
in interpreting the development and evolution of dark matter
clustering (e.g., Davis et al.~1985; Colin, Carlberg, \& Couchman 1997;
and  Jenkins et al.~1998).
The modeling of the connection between the clustering of galaxies (which
are the most easily observed component) and dark matter (which provides
the gravitational field), however, is complex and much less well
understood.
In essence, deriving such a connection requires a full understanding of 
galaxy formation and evolution.

On the observational side, progressively larger redshift surveys
of galaxies have provided relatively robust measurements of the
present-epoch galaxy correlation function or power spectrum
 (e.g., the CfA survey [Davis \& Peebles 1983, Vogeley et al. 1992]; 
and the LCRS [Lin et al. 1996]).
Although there has been a number of investigations attempting
to measure the evolution of galaxy clustering out to $z\sim$ 0.5 to 1.0, 
they were all based on
small surveys which were not specifically
designed for this purpose (e.g., LeF\`evre et al. 1996,
Shepherd et al. 1997, Carlberg et al. 1997).
These surveys cover very small areas with sample sizes of 100's
of objects.
Furthermore, the substantial galaxy population evolution over this
redshift range must be taken into account to ensure that similar samples of
galaxies at different redshifts are being used to measure
the clustering evolution.

The second Canadian Network for Observational Cosmology (CNOC2)
Redshift Survey is the first large redshift survey of faint field
galaxies carried out with the explicit goal of investigating
the evolution of clustering of galaxies.
Such an investigation requires a sample with a large number of 
galaxies spanning a significant redshift range and covering a 
sufficiently large area on the sky.
The redshift range over which the evolution is to be measured
is 0.1 to 0.6, chosen
to maximize the efficiency of a 4m class telescope.
The survey is also designed specifically to
provide a large database of both spectroscopy and
multicolor photometry data for the study of 
the evolution of galaxy populations over the same epochs.
This is particularly important as such information will allow us to
attempt to
disentangle any inter-dependence between the correlation function 
and galaxy types and evolution.

The current knowledge of field galaxy evolution at $z\sim0.5$ is
based mostly on the measurement of the luminosity function (LF) (e.g.,
Lilly et al.~1995; Ellis et al.~1996; Lin et al.~1997), with the
primary conclusion being that the most rapid evolution is seen
in blue, star-forming galaxies.
A much larger redshift sample of several thousand galaxies with
multi-color photometry will vastly improve on the current results.
The multicolor data can provide sufficient information to allow
a classification of the spectral energy distributions 
of the sample galaxies. 
This, along with a large sample size, will make possible a much more 
detailed analysis of the evolution of the LF as a function of galaxy 
population, perhaps enabling a definitive discrimination between 
luminosity and density evolution.

In this paper, we describe the general strategy and design of the
CNOC2 survey, the data reduction methods, and the creation of
weight functions. 
Also presented is the first data catalog from the survey.
In Section 2 we present the survey strategy, the field selections
and the observations.
Sections 3 and 4 describe the data reduction methods for the photometric
and spectroscopic data, respectively.
The issues of completeness and weights are discussed in Section 5.
Finally, Section 6 presents the catalog for the patch CNOC$\,$0223+00.
First results on galaxy population evolution and clustering
evolution are presented in papers by Lin et al.~(1999) and
Carlberg et al.~(2000), respectively.
Reports on close-pair merger evolution and the 
serendipitous active galactic nuclei sample from the survey are
given in papers by Patton et al.~(2000) and Hall et al.~(2000).
Additional papers on more detailed studies of galaxy evolution, the
dependence of clustering evolution on galaxy types, and other 
related subjects are in preparation.

\section{The Survey}

The CNOC2 Field Galaxy Redshift Survey was conducted using 
the MOS arm of the MOS/SIS
spectrograph (LeF\`evre et al. 1994) at the 3.6m  Canada-France-Hawaii
Telescope (CFHT).
The primary goal of the survey is to obtain a large, well-defined
sample of galaxies with high quality spectroscopic and photometric
data for the purpose of studying the evolution of galaxy clustering
in the redshift range  of $\sim0.1$ to 0.6.
To first order, such a goal requires a survey at
these intermediate redshifts which is comparable
to the local universe CfA redshift survey (Huchra et al. 1983, Geller 
\& Huchra 1989) in the number of galaxies ($\sim 10^4$),
 the volume covered ($\sim10^6h^{-3}$ Mpc$^3$), and velocity
accuracy ($\sim 50$ km s$^{-1}$).

The redshift range for the survey is chosen to maximize the spectroscopic
efficiency of a 4m class telescope and the MOS spectrograph.
Based on the experience from the CNOC1 Cluster Redshift Survey
(Yee, Ellingson, \& Carlberg 1996; hereafter YEC), 
a relatively high success rate of $\sim85$\% can be 
obtained with a reasonable exposure time of the order
of an hour for galaxies at $R_c\sim21.5$.
Such a sample would have an average galaxy density on the sky of
$\sim $ 145 galaxies per MOS field (of about 70 square arcminutes), 
and a peak of $z\sim0.35$ in the redshift distribution, well-matched
to the number of slits available in the MOS field and the redshift
range.

The robust measurement of the spectral energy distributions (SED)
of the sample galaxies in the form of multi-color photometry is also an
integral part of the survey.
Furthermore, the availability of  multicolor photometry will allow 
us to produce 
a photometric-redshift training set of unprecedented size, and also
provide important consistency checks on the spectroscopic redshift 
measurements.
Hence, a significant amount of telescope
 time is also devoted to imaging in the survey.
In the following subsections, we describe the main features of the
survey in detail.

\subsection {Field Selections}

There are several considerations in choosing the survey areas on the sky.
To avoid being dominated by a small number of large structures,
the survey covers four widely separated regions, called {\it patches},
 on the sky.
There are a number of advantages in splitting the sample area into
4 regions.
First, this allows one to obtain a reasonable sampling of the clustering;
this is particularly important because each of the patches still covers
a relatively small area on the sky.
Second, having 4 patches provides a rough estimate of the cosmic 
variance in the determination of the clustering and statistical
properties of galaxies such as the luminosity function,
luminosity density, and star formation rate.
In the case of clustering, the different patches will also
provide an indication as to whether there are
significant clustering signals at scales larger than 
the area covered by a single patch.
Finally, having 4 patches allows us to distribute them
in Right Ascension to maximize observing
efficiency.

For a sample of the order of $\sim 10^4$ galaxies over a volume of
$\sim10^{6}h^{-3}$ Mpc$^3$, we need to cover the order of
at least 1.5 square degrees.
Hence, each patch should have an area of about 0.4 square degrees.
One pointing of the MOS field, after considerations of spectral
coverage on the CCD, provides a spectroscopically 
defined area of about $9'\times7'$, with about 15$''$ overlapping
area with the adjacent fields.
Thus, each patch is equivalent to about 20 MOS fields.

To obtain a sufficiently large length scale for determining 
the correlation function reliably, we also need each patch to extend about 
10 to 20 \hinv Mpc over our redshift range.
These considerations motivated us to design the geometric shape of
each patch to have a central block of about 0.5$^{\rm o}$$\times$0.5$^{\rm o}$ 
with two orthogonal ``legs'' extending outward, as illustrated in Figure 1.
Each patch, as designed, spans $\sim80'$ 
and $\sim63'$ in the North-South and East-West  direction, respectively.
We have named the fields in each patch 
as indicated in Figure 1 (along with the sequential field numbers), 
with the {\bf a} fields representing the North
arm, the {\bf b} fields representing the East portion of the central block,
and the {\bf c} fields representing the West edge and arm.
The somewhat peculiar naming of the {\bf b} fields is due to the initial
design of the patch having an inverted $Y$ shape, and the name {\bf b} was
used to designate the East arm.
The patch design was altered after the first run in order to include
a central block, so that the {\bf b} arm has effectively ``curled'' into
the middle to form the central block.

\begin{figure}[ht] \figurenum{1}
\plotone{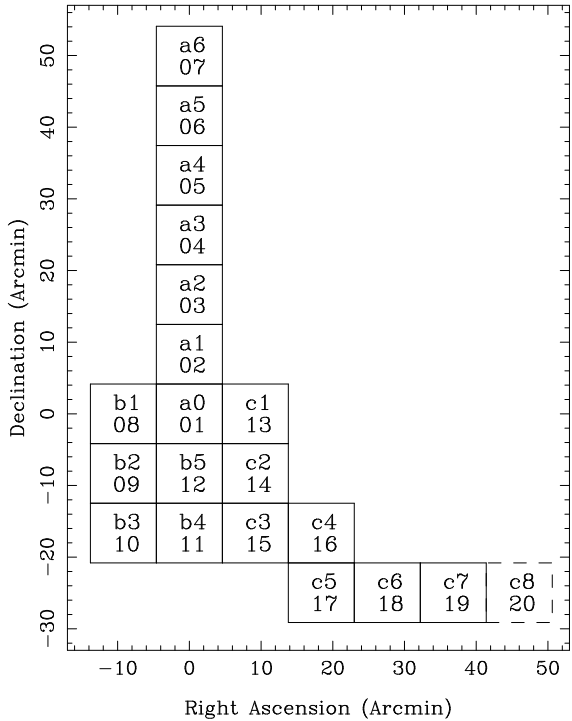} 
\caption{\footnotesize
Geometric shape of a patch and the designations of
fields. North-East is upper left.  The field size is based
on that for the STIS2 CCD; fields taken with the ORBIT1 and
LORAL3 CCDs are slightly smaller in the Declination direction.
Although the {\bf c8} field was included in the original design,
none of the patches actually contains one.
}   
\end{figure}

The properties of the patches are listed in Table 1.
The four patches are selected based on several criteria.
They are chosen to avoid bright stars ($m\ltapr 12$), low-redshift
clusters (e.g., Abell clusters) and other known low-redshift bright galaxies,
and known quasars and AGN.
The patches have Galactic latitudes between 45$^{\rm o}$ and 60$^{\rm o}$.
The lower bound guards against excessive Galactic extinction ($<0.15$
mag in $A_B$).
The extinction in the direction of each patch is obtained from
the dust maps of Schlegel, Finkbeiner, \& Davis (1998) and 
listed in Table 1 (see Lin et al. 1999 for details).
The upper bound is chosen to ensure that there are a sufficient number
of stars in each MOS field to provide proper star-galaxy classification.
A significant number of stars are needed because of the variable
point-spread function (PSF) over the MOS field (see Section 3.2).

The positions of the patches are chosen so that they are
5 to 7 hours apart in Right Ascension.
This allows two patches to be accessible on the sky at any time of
the year, and ensures that observations can be made within reasonable
hour angles at any part of a given night.
A total of 74 fields out of the intended 80 were observed for the survey,
meeting 92.5\%~of the initial design goals.

\subsection {Observations}

We use the observational techniques developed for the CNOC1 
Cluster Redshift Survey (YEC), with some additional improvements
to increase the efficiency and in the sample selection method.
The survey was carried out at the CFHT 3.6m telescope using the
MOS  imaging spectrograph.
The survey was completed over a
total of 53 nights in 7 runs from February 1995 to May 1998,
 with approximately 32 usable nights.
Three different CCDs were used over the lifetime of the survey.
Some pertinent properties of the CCDs are listed in Table 2 and
a journal of the observations is presented in Table 3.

Three CCDs were used due to various reasons.
Initially, the ORBIT1 CCD, a high quantum efficiency (QE) and blue
sensitive detector, was used for run no. 1.
However, it died just before run no. 2, and the older LORAL3, which
has lower QE and poor blue sensitivity, was pressed back into service
for both runs no. 2 and 3.
From runs no. 4 to 7, the new STIS2 CCD became available.
This is a high QE and blue sensitive detector with very clean
cosmetic characteristics, but with a larger pixel size.
Although the STIS2 CCD is a 2048$\times$2048 detector, the total
size is significantly larger than the available field of view of
MOS, and hence only a portion of the CCD was used.
We note that the majority of the data (78\%)
were obtained using the STIS2 CCD, as the first 3 runs
were plagued by bad weather. 
The exposure times for both spectroscopic and direct images
were adjusted to compensate for the different QEs of the CCDs.

The MOS spectrograph has a usable field of about 10$'$ diameter
with the corners being in poor focus.
An area smaller than the total imaging area is defined as 
the spectroscopic field, which is the area used for the survey.
This defined spectroscopic area is limited by the
CCD detector's extent in spectral coverage, in that slits placed
on different parts of the chip must all produce a complete spectrum.
The spectroscopic field size for each detector is listed in Table 2.
The defined area of adjacent fields are nominally overlapped 
10$''$ to 20$''$ to provide consistency checks on astrometry, 
photometry, and redshift determination.
Note that the STIS2 CCD has a significantly larger defined area
because of the larger physical CCD size allowing for additional
areas for the spectra of objects at the edge of the imaging field.

As in the CNOC1 survey, we use a band-limiting filter to increase
significantly the multiplexing efficiency of the survey.
The shortened spectra allow for multi-tiering of spectra on the
CCD image, increasing the number of slits per spectroscopic mask
from the order of 30 to about 100.
For the CNOC2 survey, we have designed a filter with band limits
of $\sim$ 4300\AA~to 6300\AA.
These limits were chosen with various compromises in mind.
The range must be short enough to allow for significant multi-tiering,
but long enough to be able to sample the key galaxy spectral features
over a redshift range commanded by an apparent magnitude limit which 
is optimal for a 4m class telescope.
Furthermore, the filter limits were chosen to coincide with
the onset of the grism/CCD inefficiency in the blue, and
with the first prominent atmospheric OH emission complex in the red.
The transmission curve of the band-limiting filter is shown in Figure 2.
The half-power limits for the curve are 4387\AA~and 6285\AA~with
a peak transmission of $\sim 0.8$.

\begin{figure}[htb] 
\figurenum{2}
\plotone{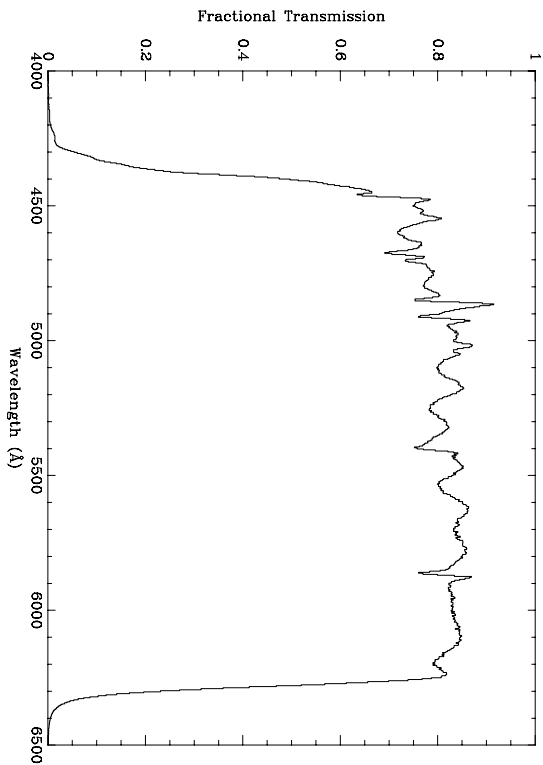}
\caption{\footnotesize
Transmission curve for the band-limiting filter for
spectroscopy measured by observing a standard star.
}
\end{figure}

The band limits effectively define the redshift completeness boundaries
of the survey.
The most prominent spectral features in galaxies over these wavelengths
are [OII]$\lambda3727$ for late-type galaxies and 
the Ca$\,$II H K features at $\lambda\lambda$3933,3969\AA~for early-type
galaxies.
For the HK lines, the effective redshift range is about 0.12 to 0.55.
Although the [OII] line disappears at the blue edge at $z\sim0.16$,
the effective low redshift boundary for emission-line galaxies is $z\sim0$,
since for most emission-line objects, detectable 
[OIII]$\lambda\lambda$4959,5007\AA~move into the red end of the spectrum
at $z\sim0.25$.
Hence, the effective redshift range for the full sample is 0.12 to 0.55;
whereas it is 0.0 to 0.68 for emission line galaxies.

The imaging observations are carried out using 5 filters: $I_c$, $R_c$,
$V$, $B$, and $U$, with typical integration times for each CCD
listed in Table 2.
(For simplicity, the subscript denoting the Cousins $R$ and $I$ filters
will be dropped for the remainder of the paper.)
The $V$ images are actually obtained through a Gunn $g$ filter,
but calibrated to the Johnson $V$ system (see Section 3.1).
The $R$ and $B$ images are utilized for designing the masks used for
the spectroscopic observations.
Image quality ranges from 0.7$''$ FWHM to 1.2$''$ in the best
focused part of the $R$ images,
with a deterioration up to about 20\% near the edges due to 
the focal reducer optics.

The imaging data are reduced at the telescope, and preliminary catalogs
of each field are produced and used to design the masks for multi-object
spectroscopy.
The mask design procedure, using a computerized algorithm,
is described in Section 2.3.
The MOS/SIS system at CFHT has a computer controlled laser  slit
cutting machine (LAMA), which allows spectroscopic masks with as many
as 100 slits to be fabricated in 20 to 40 minutes.
The elapsed time between obtaining a set of direct images to
having a mask for spectroscopic observation, including the procedures
for preprocessing, catalog creation, mask design, and mask cutting,
was pushed to as low as about 2 hours by the end of the survey.

  The new B300 grism is used for spectroscopy.
This grism has enhanced blue response compared to the
O300 grism used for CNOC1. 
Along with the blue sensitive CCDs, this produces a significant
improvement in the blue region of CNOC2 spectra.
The grism is blazed at 4584\AA, and has a dispersion
of 233.6\AA/mm, giving 3.55\AA/pixel for the ORBIT1 and 
LORAL3 CCDs, and 4.96\AA/pixel for the STIS2 CCD.
A slit width of 1.3$''$ is used for the spectroscopic observation,
giving a spectral resolution of $\sim$14.8\AA.

The observing procedure is similar to that for the CNOC1 survey (YEC).
Two masks are observed for each field, with nominal integration times
differing by a factor of $\sim2$ between the A and B masks.
Table 2 lists the spectroscopic integration times used for each CCD.
Because the need to have continuously available a set of masks
for spectroscopic observations prepared in advance,
to maximize the observation efficiency, a computered-aided schedule
of the imaging and spectroscopic observation sequence for each night
is prepared ahead of time, and adjusted as weather, equipment failure,
and varying acquisition and set-up time required.
With the STIS2 CCD which has higher QE and faster readout time, 
on average for each field, including overhead for acquisition,
focusing, mask alignment, and arc lamp calibration,
$\sim$155 minutes were need for the spectroscopic
observation of two masks, allowing as many as 800 spectra to be obtained
in a single 10-hour clear night.
For the direct imaging, 
$\sim$60 minutes were required for each field. 

\subsection {Galaxy Sample and Mask Design}

A computerized mask design algorithm is used for generating the
positional information of the slits for the fabrication of masks.
Besides allowing one to optimize the placement of as many as 100
slits per mask, an objective design algorithm based on  
properly calibrated photometric catalogs also serves the very
important task of producing a well-understood and well-defined
spectroscopic galaxy redshift sample.
The algorithm for optimal slit placement is identical to that used for
CNOC1 (YEC); however, the selection process for the galaxy sample is
different, and is specifically designed to generate a fair field-galaxy
sample based on both the $R$ and $B$ photometry.
The mask design uses only objects classified as galaxies 
with occasional stars included, either serendipitously (e.g., falling
into a slit designed for a galaxy), or due to misclassification (see
Sections 3.1 and 3.2).

Two masks, A and B, are designed for each field.
For 4 of the 74 fields, a third mask C is also observed when either 
mask A or B are deemed insufficient due to poor observing conditions.
The total number of masks observed in each patch is listed in Table 2. 
The nominal spectroscopic limit is $R=21.5$ and $B=22.5$.
The defined sample for the spectroscopic survey
is the union set of the two limits.
Objects fainter than both of these limits belong to the secondary 
sample, for which slits are placed only if sufficient room for the 
spectrum is available on the detector.

The two-mask strategy for multi-object spectroscopy offers three
important advantages.
First, by choosing objects with different average brightnesses 
for the two masks,
the integration time for each mask can be geared to the brightness of
the objects, resulting in a significant saving of exposure time.
Second, the second mask allows for the compensation of under-representation
of close pairs in selecting objects for spectroscopic observation.
This under-representation arises from the fact that
once a slit is placed, a certain
area on the CCD is blocked by the resulting spectrum of the object.
This blockage as a function of the distance from a chosen object
 is illustrated by the thin solid line on Figure 3 which is
derived based on the area blocked by the spectrum as a function of
radius from the center of the slit.
The discontinuity arises from the spectral coverage not being 
symmetric with respect to the central wavelength.
This function is equivalent to the pair fraction expected in a
spectroscopic sample created by a single mask from a parent
photometric sample distributed uniformly on the sky.
The actual situation is worse in that galaxies are clustered on the sky.
The use of the second mask allows objects being blocked by the
spectra in the first mask to be observed.
This is a particularly important feature, in that without such
compensation,  a fair sampling of object separations is not possible,
rendering severe selection effects in the correlation function.
Finally, a second mask allows for the redundant observations of 
a significant number of objects, which is important for both quality
control and obtaining empirical estimates of redshift
measurement uncertainties.

\begin{figure}[ht] 
\figurenum{3}
\plotone{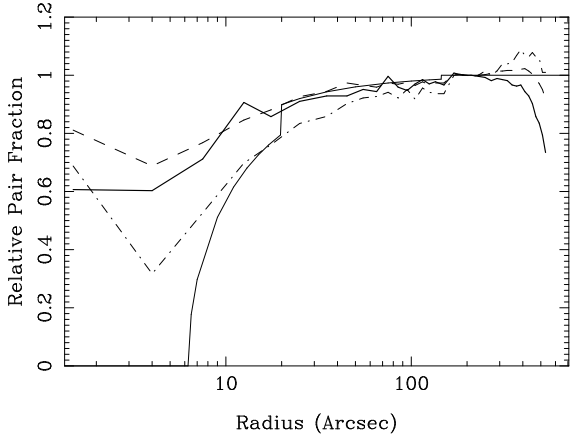} 
\caption{\footnotesize
Pair distributions for galaxies with $R_c\le21.5$ mag 
as a function of separation for the whole survey.
The pairs are counted in individual fields and summed. This
provides an indication of the effect of the mask design algorithm.
Each distribution is plotted in pair fraction relative to the
total photometric pair distribution, normalized to 1.0 at the
radius of 220 arcsecond.
The dot-dashed line represents the distribution of the A mask
sample, while the dashed line shows the distribution of the
total slit sample (i.e., both A and B masks)
 selected by the mask design algorithm.
The light solid line is the theoretical distribution computed
based on the fraction of area blocked by a single spectrum as a
function of radius from the center of the slit.
It represents the expected pair distribution arising from a
uniformly distributed sample of galaxies selected using a single
mask.
The thick solid line represent the pair fraction distribution
of the redshift sample.
See the text for a detailed discussion of these distributions.
}

\end{figure}

Mask A is designed with the emphasis on brighter galaxies, and has a
total integration time of about 1/2 of that for Mask B.
The design of mask A is based strictly on the apparent
magnitude of the galaxies.
A priority list of galaxies brighter than $R=20.25$ mag is made,
with the objects closer to $R=20.25$ mag having the higher priorities.
Slits which produce non-overlapping spectra are placed based 
on the order from this list.

This simple algorithm, however, produces the undesirable effect that
objects near the upper and lower edges in the dispersion direction
(i.e., the north and south edges)
will preferentially have a higher probability of being placed.
This arises because the spectra of these objects will extend beyond
the defined field area on the CCD, effectively reducing the
probability of their spectrum being blocked by those belonging to
objects already chosen.
To compensate for this edge effect, whenever a galaxy is chosen,
the fraction of its spectrum falling outside the defined field
is computed.
The object is not accepted, and placed at the end of the priority
queue, if a random number drawn between 0 and 1 is smaller than this 
fraction.
This edge-effect correction is applied to all subsamples 
in the design process.
Note that this additional step does not prevent a larger fraction
of objects near the edges being selected, simply for the reason
that there is more area there to place spectra.
However, this procedure ensures that the selection of slits in
the remainder of the field is not unduly driven by the excess placements
of slits near the edges, since the latter objects have their priority
systematically suppressed.

Once all possible simple placements have been done,
the number of placements is optimized by shifting existing slits along
the spatial direction (i.e., so that the object may no longer be in the
center of its slit) in order to place additional slits (for details,
see YEC).
When no more slits from this sample can be placed, a secondary
sample is produced from the remainder
of the photometric sample based on an ordered list of increasing
magnitude starting from $R=20.25$, and the whole procedure is repeated.

The design goals for the B masks are considerably more complex.
The B mask is designed to complement the A mask in both the magnitude
priority and close-pair selection.
The primary sample for the B mask 
contains objects  not already placed in the A mask with 
$R=20.25$ as the lower bound and $R=21.50$ or $B=22.5$ as the upper bound.
These objects are ordered in increasing $R$ magnitude.
Note that this produces
an effect that bluer objects of similar $B$ magnitudes
will be slightly undersampled compared to the redder objects.
However, the net effect is expected to be slight, since faint
blue galaxies typically have emission-line spectra which are
more likely to yield measured redshifts.

A second ranking, designed to compensate for the close pair selection
effect, is produced in the following manner.
First, the ratios of the number of pairs as a function of separation
for both the objects already assigned
to the A mask and in the $R<21.5$ sample are derived.
For each object in the sample that is not already in mask A, the
distance to the nearest object that is already assigned to mask A
is determined.
The ratio of the pair distributions at the appropriate separation
is then used to prioritize the pair-compensation ranking, in that
the lower the ratio, the higher the ranking.
The final priority ranking of the object is then created by adding the
magnitude and pair-compensation rankings in quadrature.
Slits are placed on the mask based on these rankings.
And again, additional target placements produced by shifting the 
existing slits are made to maximize the number.

The pair-compensation applied to mask B, however, leaves one possible
selection effect.
The objects fainter than the mask A primary magnitude limit ($R=20.25$)
which are placed on mask A will have a shorter exposure time than those
in the same magnitude range in mask B.
Hence, even though both objects of a close pair may be chosen
(i.e., one in mask A and the other in B),
one of these will have a lower chance of having the redshift measured.
This problem is partially alleviated by the slit position 
shifting algorithm to add additional objects (on the same mask),
which is also done with the pair-compensation priorities.
The slit-shifting procedure has the effect of freeing up some of the 
blocked areas.

Once the primary sample for mask B is exhausted, a second sample, consisting
of all objects with $R<20.25$ that have not been placed in mask A
(i.e., those in the primary sample of A that were missed ), is
created, and the same procedure (with the pair-compensation
ranking) is applied, with the exception that
the magnitude ranking is done from faint ($R=20.25$) to bright.
The third and fourth samples are redundant observations of objects
already placed in mask A: galaxies with $R>20.25$ and $R<21.5$ or
$B<22.5$, and galaxies with $R<20.25$, respectively.
Finally, any additional available space is filled with the final
sample of galaxies with $R>21.5$, ranked by brightness.

\begin{figure}[ht] 
\figurenum{4}
\plotone{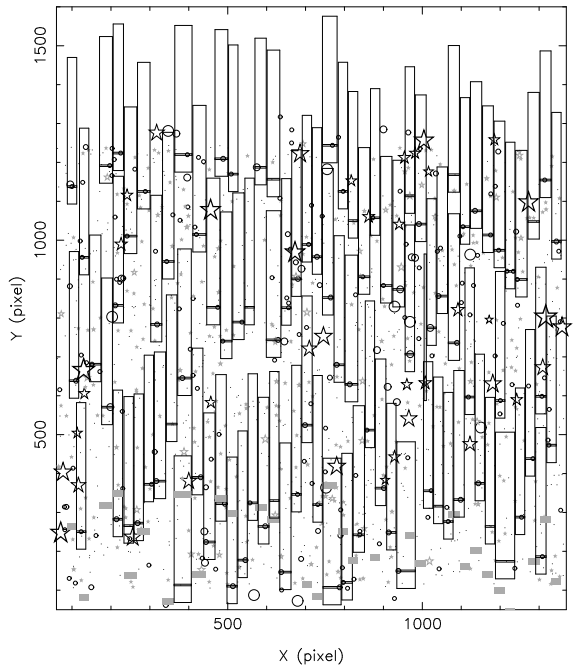} 
\caption{\footnotesize 
A schematic map of a typical mask design: mask B for
the field CNOC$\,$0223+00{\bf b5}.
Objects from the photometric catalog are denoted by circles (galaxies)
and stars (stars) with their sizes indicating relative magnitudes.
Slit positions are marked with small horizontal rectangles, while
the vertical boxes indicate the positions of the spectra.
The solid gray boxes mark the positions of the zeroth order
contamination.  There are a total of 95 slits on this mask.
}
\end{figure}

A typical mask design is shown in Figure 4.
In Figures 5 and 6, we show the magnitude distribution and fraction 
of objects selected as a function of magnitude
for masks A and B for the fields in the Patch CNOC$\,$0223+00.
Note that bright galaxies have a higher sampling rate relative
to the fainter galaxies.
This is part of the design philosophy, as the smaller total number
of bright galaxies requires a higher sampling rate to provide
sufficient statistics.

Figure 3 also presents the pair fraction as a function of separation
for objects in the A masks (dot-dashed line) and in 
the A and B masks (dashed line) for the whole survey,
summed field by field,
demonstrating the corrective action of having two masks.
Note that the two-mask and pair-compensation procedures are still not
sufficient to completely correct the lower sampling rate at
separations up to about 100$''$.
This is simply due to the fact that with only two masks, high
galaxy density regions will always be undersampled.
Furthermore, at small angles,
close triples will be severely undersampled in a two-mask system.
The better-than-expected coverage at the smallest angular bin ($<3''$)
is due to serendipitous observations of 
very close doubles in a single slit.
At large separations ($>300''$), the pair fraction is oversampled.
This is due to the edge effect of objects near the North/South edges
having a higher probability of being placed.
Also plotted in Figure 3 (thick solid line)
is the pair distribution of the final redshift
sample, which follows the mask pair distribution except at large
radii, showing a significant drop despite more objects being
sampled by the masks at these separations.
This drop is due to the poorer image quality at the edges and corners
of MOS.
The undersampling effect at the arcminute scale
is partially corrected by applying a geometric
weight (see Section 5) to each object.
Nevertheless the redshift sample pair distribution from an entire 
patch shows periodic spatial features at an amplitude of about
10 to 20\% with a period of about the width of the fields($\sim550''$)
 in the East-West direction.
These features are noticeable in the E-W direction but not in the N-S
direction due to the fact that extra objects are sampled by the masks
at the North/South edges, but not the East/West edges.

\begin{figure}[ht] 
\figurenum{5}
\plotone{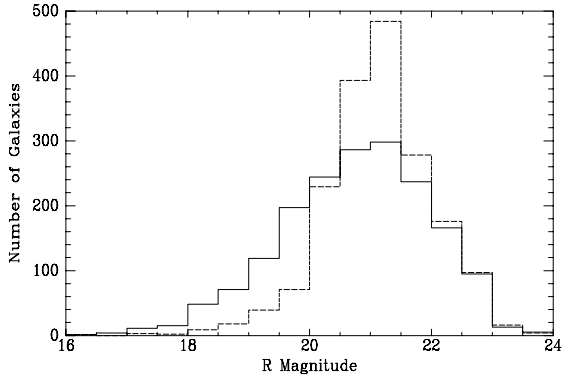} 
\caption{\footnotesize
The magnitude distribution for objects selected by the mask design algorithm
for the A masks (solid histogram) and B masks (dashed histogram) for
Patch CNOC 0223+00.
}
\end{figure}
\begin{figure}[ht] 
\figurenum{6}
\plotone{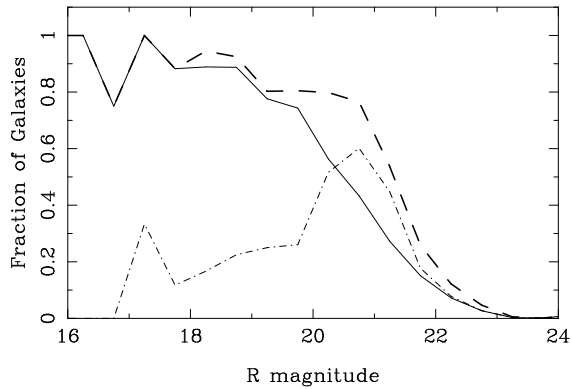} 
\caption{\footnotesize
The fraction of objects as a function of magnitude 
selected by the mask design algorithm
for the A masks (solid line), B masks (dot-dashed line) and either mask
(heavy dashed line) for the fields in Patch CNOC 0223+00.
The fraction for either mask is not identical to the sum of masks A and
B due to redundant selections.
}
\end{figure}

\section{Photometric Data Reduction} 

\subsection{Photometry}

  Photometric reduction is performed using the program PPP
(Yee, 1991) as described in YEC.
Detailed descriptions of object finding, star-galaxy classification,
and ``total'' photometry via growth curve analysis can be found
in YEC and Yee (1991).

The photometric catalog in 5 filters is produced using a master
object list created from the $R$ image.
Although in principle, this procedure will miss extremely blue
objects, or red objects in $R-I$;
in practice, such an effect is not expected to be significant,
since the $R$ band image is by far the deepest in the set.
Visual inspection of the images from the other
filters indicates that at worst only the occasional very faint blue
object is missed, and the $R$-selected catalog (which is
100\% complete up to about 23.2 mag) should be complete for
objects redder than the extreme color of $B-R=-0.7$ at 
the $B=22.5$ limit of the primary spectroscopic sample.
We note that the photometric catalog contains no objects
as blue as $B-R=0.0$, well above the cut-off limit.
However, it is clear that these catalogs should not be used to
search for $R$-band drop-outs!

The object list is visually inspected, and spurious objects arising
from cosmetic defects such as bleeding columns and diffraction spikes,
and small structures (e.g., HII regions) within large, well-resolved, 
low-redshift galaxies are removed.
We also pay close attention to close doubles that are missed by
the object finding routine, and manually add these to the list.
This adjustment is important in preventing stars with a close
companion from being misclassified as galaxies.

Because of the slightly different distortion from the optics
of the focal reducer of MOS over the large wavelength region
covered by the 5 filters, and the fact that three different CCDs
with different scales have been used, the master list of objects 
from the $R$ image is corrected to the co-ordinate system of the other
filters via a geometric transformation determined by comparing
positions of bright objects in the images.

The photometry from the longer exposure $U$ band images suffers
significantly more severely from cosmic-ray hits than the
other bands.
This is due to both the longer exposure time and the much lower relative
flux of the desired signals.
Hence, a cosmic-ray detection removal algorithm (as part of the
PPP package) is applied
to all the $U$ band images before photometry is done.
This algorithm detects cosmic-ray hits by first identifying
all pixels that are more than 9$\sigma$ above the local sky
root-mean-squared fluctuation.
These pixels are then tested to see if they are consistent with
being part of a real object via a sharpness test.
If they are not, they are tagged as cosmic ray detections, and
a padding of an additional layer of pixels around them is applied
to mask any residual lower level extension of the detection.
This simple algorithm works extremely well, and does not accidentally
tag any real object pixels.
The tagged pixels are then ``fixed'' by simple linear interpolation.
It is found that by applying this algorithm, the $U$ band photometry
improves for about 5\% of the galaxies based on improved
SED fits for objects in the redshift sample.

We note that the photometry catalogs 
created at the summit in real time (in $R$ and $B$ only) which
are used for designing the masks are preliminary.
These catalogs are updated later with a more careful inspection and
adjustment of the star-galaxy classification parameters.
Hence, there are small improvements in the final catalogs compared
to the summit catalogs.
In general, the final catalogs feature fewer spurious objects, 
fewer missed closed neighbors, and more reliable
the star-galaxy classifications (see Section 3.2),
especially for objects situated
near edges and corners of the image where the defocusing is significant.
However, because the mask design uses the summit catalogs, a number
of stars misclassified as galaxies are unwittingly selected for 
spectroscopic observation.
Typically these stars are either near corners or edges, or have a faint,
close neighbor.
Some of these misclassified objects are active galaxies
and quasars; they are discussed in Hall et al. (2000).

The calibration to standard systems is achieved using three to four
Landolt (1994) standard fields, plus M67 whenever possible, 
per each run in the 5 filters.
It is found that the calibration of the $g$ band photometry requires
a rather large and uncertain color term.
This is due to the fact that the $g$ filter used is actually significantly
redder than the original $g$ band definition (Thuan \& Gunn 1976),
and that there are only a small number of standard stars available
(compared to the Landolt system).
Hence, we have calibrated these images to the  $V$ system based
on Landolt standards.
This results in a smaller color term and in general 
more stable results.
Nevertheless, for completeness we also produce catalogs with the red
and the green filters calibrated to the Gunn $r$ and $g$ system.
Typical systematic uncertainties in the zero point calibration
constants are 0.04, 0.04, 0.05, 0.05, 0.07 for $IRVBU$.
Most of the data are obtained under photometric conditions.
Small adjustments to compensate for field-to-field differences are made 
using newly obtained large format CCD images (KPNO 0.9m MOSAIC camera and
CFH12K camera) and using the overlapping regions between adjacent fields.
While this does not eliminate the systematic uncertainties, it does put
every field in a single patch on a consistent calibration.

The final ``total'' $R$ magnitude for each object is created using
the following algorithm.
The magnitude from the adopted optimal aperture determined
 from the $R$-band growth
curve (see Yee 1991) is used as the primary magnitude for an object.
If the aperture is smaller than $12''$,
the magnitude is extrapolated to the $12''$
aperture by using the stellar PSF shape.
This nominally produces a correction of less than 0.05 mag.
This procedure provides an exact correction for faint stars, and a first 
order correction for faint galaxies to compensate for seeing smearing.
For galaxies brighter than about 19 mag,
a second pass for the growth curve is performed using a maximum
aperture size of $24''$.
The few very bright galaxies ($R\ltapr13.5$) are typically larger than
this aperture and their magnitudes can be underestimated by 0.1 to 0.2
magnitudes.
The x-y position and star-galaxy classification for each object
 are also adopted from the $R$ image.

To derive the object magnitude for another filter, we
use the color of the object relative to a reference filter.
A default color aperture,
which is the largest aperture used for color determination between
a pair of filters, is set at a relatively large value of 4$''$
to account for the non-uniform PSF in the MOS image.
The color between the two filters is formed using the flux
inside a color aperture which is the minimum of the
adopted optimal apertures for the two filters and the
default color aperture.
Note that the adopted  color aperture radius in general is not a specific
aperture used in the tabulated growth curve (which uses apertures
with integral pixel diameters).
The magnitudes within the color aperture are derived by interpolation
of the quantized growth curve.
The total magnitude for the filter in question is then the difference 
between the optimal magnitude for the reference filter, 
which is nominally chosen to be $R$, and the color.
For the $U$ band, the color is formed using $B$ and $U$ to avoid 
using the long baseline between $R$ and $U$ which may increase the uncertainty.
This method of determining the ``total magnitude''
in each filter has the advantage of always having colors defined
from the same aperture for the two filters, and also decreasing
the uncertainty in the flux measurement of the fainter filter.
However, it does make the assumption that there is a negligible
color gradient in the object.
The photometry uncertainties for each filter in the color pair
at the adopted color aperture are recorded, and the uncertainty
of the color is the quadrature sum of the two.

The positions of objects on the CCD frame are determined using an
iterative intensity centroid method (see Yee 1991).
The uncertainty ranges from better than 0.01 pixels for bright
objects to about a pixel for objects near the 5$\sigma$ detection limit.
However, the major uncertainty in the relative positions of objects
from the same field arises from the pincushion distortion
of the focal-reduced image, which could be as large as 5 to 6 
pixels at the corners.
The distortion is corrected using an image taken through a mask with a
grid of pinholes of known separations.
In general, the relative positions of objects near the edges and corners 
may be uncertain systematically up to 1$''$.
The star cluster M67 is used as the astrometry reference 
(Girardi et al. 1989) for determining the scale and rotation of 
the CCD set-up, to an accuracy of about 0.0004 and 0.05$^{\rm o}$,
respectively.
Note that these uncertainties compound the inaccuracy of
the  absolute position determinations
of objects over the large distances spanned by the fields.

\subsection{Star-Galaxy Classification}

Star-galaxy classification for the MOS imaging data is particularly
challenging due to the large variation in the image quality due to the
focal reducer optics.
The focus variation across the field is often
sufficient to produce crescent- or even donut-shaped
PSFs at the corners.
We adopt the observational procedure of performing focusing 
consistently at the same predetermined region about 1/3 of the
way outwards from the center of the CCD, so that
within the defined area the image quality variation is minimized.
A variable PSF classification scheme, as outlined in YEC, is used.
This procedure essentially compares the growth curve shape of
each object with the four nearest PSF standards, one in each
quadrant centered on the object (see Yee 1991 and YEC).
The PSF standards are chosen automatically; however, 
manual intervention in the choices of PSFs 
is often required near the corners and edges of the image.
Typically 20 to 40 reference PSFs are used per field.

Stars having a very close neighbor of either a galaxy or another
star are occasionally misclassified as non-stellar.
These misclassifications are corrected by hand during visual
inspection of the classifier plot.
In general, objects down to a brightness of $R\sim22$ mag have
robust star-galaxy classification, beyond which some faint galaxies
begin to merge into the stellar sequence (e.g., see Figure 2 in YEC).
A statistical variable star-galaxy classification criterion is used
for the faint objects (see Yee 1991) to compensate for the merging
of the star and galaxy sequences in the classifier space.
While the separation between stars and galaxies is excellent
at magnitudes brighter than the nominal spectroscopic limit of $R=21.5$,
quasars, luminous distant active galactic nuclei, and a
small number of compact galaxies may be missed in the spectroscopic
sample.
A more detailed discussion of the effect of star-galaxy classification
on compact extragalactic objects is provided in
Hall et al.~(2000) which analyses the sample of serendipitous AGN and
quasars in the CNOC2 sample,

\section{Spectroscopic Reduction} 

\subsection{Spectral Extraction}

The spectrum of each object is extracted and wavelength and flux calibrated
using semiautomated IRAF\footnote {
The Image Reduction and Analysis Facility (IRAF)
is distributed by the National Optical Astronomy
Observatories, which is operated by AURA, Inc., 
under contract to the National Science Foundation.} reduction procedures.
Many of the procedures are the same as those used for the CNOC1 survey
as described in Section 6.2 of Yee, Ellingson \& Carlberg (1996).
Here we summarize those procedures and discuss in detail 
only those changes made for CNOC2. The scripts and programs used are
available from G. Wirth (wirth@keck.hawaii.edu).
A major conceptual change is that instead of extracting all the spectra
from the same large image, each individual spectrum is copied to a smaller 
subimage upon which the extraction is performed.

Each mask has a flatfield and arc lamp image 
and typically two individual spectroscopic exposures.
The individual spectroscopic exposures are cleaned of cosmic rays (see YEC),
overscan subtracted and summed
after applying small shifts to correct for flexure in a handful of cases.
Regions of zero-order contamination are interpolated over on the flatfield
after being marked interactively. 

Since the relative position of each spectrum on the CCD is known exactly 
in advance, for each mask it is easy to construct an automatic extraction
file 
containing the spatial and dispersion position of the objects and the
relative positions of the edges of each slit.
The automatic extraction file is used to create an IRAF aperture database
containing the aperture position, width, and sky background ranges for each
object, including serendipitous objects not targeted in the mask design.
Three subimages are then created for each object by copying the same region
from the summed spectroscopic exposure, flatfield image, and arc lamp image.
Each flatfield subimage is normalized using a cubic spline fit to the mean 
flat field wavelength profile, leaving only pixel-to-pixel variations.  
The resulting response subimage is divided into the object subimage
to yield a flattened object subimage.

Each object subimage is interactively examined and the default object and
background apertures adjusted if necessary.
Notes are made at this stage of any salient features of the spectrum,
including emission line(s), a very faint or invisible continuum,
overlapping spectra, multiple objects per slit, etc.
The spectrum is then extracted using variance weighting
and an arc spectrum is extracted from the arc lamp subimage
using the same extraction aperture and profile weighting.
The arc lamp images contain lines 
from He, Ne and Ar, and give 11 lines strong enough and sufficiently 
unblended to be used for wavelength calibration. There is a gap 
with no arc lines between 5015\AA\ and 5875\AA\, where the wavelength 
calibration is less certain.
The wavelength calibration solution for each subimage is found by 
interactively identifying several arc lines, using a line database 
to identify the rest, and then fitting a cubic polynomial to the data. 
The resulting wavelength solution is non-linear (with deviations 
of over 10\AA\ from linear). Typical rms residuals to the fit 
for the 11 arc lines 
are less than 0.1\AA. The wavelength solution changes 
significantly for spectra at differing locations on the CCD, with
the mean dispersion ranging from 4.9\AA\ per pixel at the bottom 
to 5.1\AA\ per pixel at the top (for the STIS2 
CCD).
The object spectra are wavelength calibrated,
and all linearized to run from 4390\AA\ to 6292.21\AA\ with
a uniform dispersion of 4.89\AA/pixel. The 
wavelength solutions are checked by visual inspection of the 
wavelengths of the bright sky lines.
The data are then extinction corrected and flux calibrated to $F_{\lambda}$.
The flux calibration uses observations of standard stars, generally 
taken through one
of the central apertures. Using these stars, the end-to-end throughput
(photons hitting the telescope primary, through to electrons 
detected by the CCD) for the typical seeing is  
fairly flat across the blocking filter bandpass, with measured 
values between 10 and 15\%. Slits near the corners of the mask 
suffer from significant vignetting, which has not been 
corrected for in the current data.
As with CNOC1, the relative flux calibration should be considered 
accurate to $\lesssim20$\%
across large wavelength ranges. The absolute flux calibration of the 
spectra is considerably more uncertain.

Regions 45\AA\ wide around the
bright night sky emission lines at 5577\AA\ and 5892\AA~are
 automatically interpolated over.
Finally, the spectra are examined and residual cosmic rays 
or other bad regions are interactively marked and interpolated over, 
including regions 125\AA\ wide around zero order contamination.
These interpolated spectra are the final versions used for redshift
determination, but we also found it advantageous to keep copies of the 
uninterpolated spectra to help confirm redshifts where a feature 
falls within an interpolated region.
A total of 14932 spectra were extracted
during the course of the CNOC2 data analysis.

\subsection{Cross-Correlations}

We determine redshifts using the cross-correlation technique
described by YEC ( see also Ellingson \& Yee 1994).
Our method is similar to standard techniques (e.g., Tonry \& Davis
1979), except for a second calculation of the cross-correlation 
function, which is used to remove biases resulting from the large
redshift range we need to consider ($0 < z \lesssim 0.7$), 
coupled with the finite spectral coverage of both our object and 
template spectra.
The reader is referred to YEC for details.

The object spectra are cross-correlated against three 
galaxy templates, specifically elliptical, Sbc, and Scd galaxy spectra
taken from the spectrophotometric galaxy atlas of Kennicutt (1992).
All object spectra, along with their associated cross-correlation functions 
against each template, are visually inspected to verify or reject the redshifts
first determined automatically by the cross-correlation program on the basis
of the values of the cross-correlation coefficient $R_{cor}$ 
and peak heights (Tonry \& Davis 1979).
Generally the redshift corresponding to the template with the highest
$R_{cor}$ value is adopted, but occasionally a template with a somewhat lower 
$R_{cor}$ value may be chosen if visual inspection deems 
it a better spectral match to the object galaxy.
Low signal-to-noise spectra with indeterminable or uncertain redshifts, 
typically with $R_{cor} \lesssim 3$, are rejected.
Figure 7 shows example spectra and cross-correlation functions 
for galaxies of different spectral types and different $R_{cor}$ values.
We assign the galaxy spectral type, denoted ``Scl,'' according to the 
cross-correlation template chosen, where Scl=2, 4, and 5 correspond to 
the elliptical, Sbc, and Scd templates, respectively.
The visual inspection also includes examination of the 
two-dimensional spectral images, both before and after cosmic ray removal,
which allows us to reject any remaining cosmic rays that might otherwise
masquerade as emission lines in the extracted one-dimensional spectra.
Contaminating stellar spectra, as well as unusual spectra (e.g.,
AGN or spectroscopic gravitational lens candidates, see Hall et al.~2000)
 are also identified during the visual inspection process.
Objects classified as AGN or QSO are designated as Scl=6.
In addition, objects without assigned redshifts but with spectra
of signal-to-noise ratios judged sufficient to have detected 
spectral features if they were present are flagged and denoted by Scl=88.
Spectroscopically identified stars are designated as Scl=77.
The IRAF add-on radial velocity package RVSAO (Kurtz \& Mink 1998) is used 
to aid in graphical display and visual redshift assessment of our spectra.

\subsection{Redshift Verification and Uncertainty}

The redshifts determined during the ``first pass'' visual
inspection described above are subjected to confirmation in several 
subsequent steps.
A ``second pass'' visual inspection of those spectra assigned
a first-pass redshift is made to verify the original redshift determination
and to flag problem cases for final inspection.
This is done by one of us (H.\ Lin) on the
spectra in bulk, after all the survey data has been acquired and all
first-pass inspections completed, in order to provide a
single reasonably uniform categorization of the redshift quality.
During second-pass inspection, the original first-pass 
redshifts are assigned to 
three categories: good, probable, and questionable/bad.
Questionable/bad redshifts are subjected to a final visual inspection, 
as described below. 
Note this questionable/bad category also includes cases
where a clerical mistake was made during first pass on
an otherwise obviously good redshift.
For probable redshifts, we apply an additional, more objective 
photometric-redshift (photo-$z$) verification test to confirm the redshift
or to flag the spectrum for final visual inspection.

The photo-$z$ calibration is done using the empirical polynomial fitting
method of Connolly et al.\ (1995).
Separate fits are determined for Scl=2, 4, and 5 galaxies, using 
the full 4-patch CNOC2 data set, but the calibration sample is restricted
to those galaxies with second-pass ``good'' redshifts and 
high cross-correlation $R_{cor}$ values 
($\geq 6$ for Scl=2,4; $\geq 12$ for Scl=5).
For Scl=2 and 4 objects, the fits include up to linear-order terms in 
in the $U \! BV \! RI$ magnitudes, but for Scl=5, we also include
terms up to quadratic order to improve the fit.
We take out any redshift-dependent residuals in our photo-$z$ fit by 
subtracting the median difference between photometric and spectroscopic 
redshift, in bins of width $\Delta z = 0.1$ in spectroscopic redshift.
Note that since we are interested in checking 
the consistency between photometric and spectroscopic redshifts, rather than
trying to derive photo-$z$'s for objects with no spectroscopy 
in the first place, it is entirely valid to use existing spectroscopic
redshift and spectral class (Scl) information to optimize the photo-$z$ fits.
Using the calibration sample of second-pass good redshifts, we find
that only about 10\% of these galaxies have 
$|z({\rm photometric}) - z({\rm spectroscopic})| > 0.065$, 0.09, or 0.15
for Scl=2, 4, or 5, respectively.
We then adopt these simple cuts to flag for final inspection 
those second-pass probable redshifts which show a large discrepancy 
with their respective photometric redshift estimates.

The third and final pass is then carried out by two of us
(P.\ Hall, H.\ Lin), on the following objects:
those with second-pass questionable/bad redshifts,
those with second-pass probable redshifts failing the photo-$z$ verification
test,
those with $R_{cor}<4$,
those morphologically classified as galaxies or probable galaxies but
spectroscopically classified as stars (and vice versa),
and all objects which were not initially assigned redshifts but which do have
spectra of reasonable signal-to-noise ratios (i.e., Scl=88).
Redshifts for which both inspectors agree are good (bad) are 
retained (rejected);
those for which the inspectors disagree are resolved by a tie-breaking
vote cast by a third person (H. Yee).
We also correct any obvious mistakes in the redshift at this point
(e.g., a redshift based on a [OIII]$\lambda\lambda$5007
emission line that was mistakenly or accidentally taken to be 
a [OII]$\lambda$3727 line during first pass).
As an example,
overall about 250 redshifts out of $\sim$1550 (i.e., $\sim15$\%)
from the 0223+00 patch 
are third-pass inspected and 36 of these are ultimately excluded 
from the final redshift catalog.
Thus, the final catalog contains objects that have secure redshifts
with their quality approximately ranked by the correlation $R_{cor}$ values.

Finally, we use redundant redshift measurements, typically from independent 
pairs of A and B mask spectra, to empirically calibrate
the redshift errors calculated by our cross-correlation program.
Specifically, we examine the distribution of 
velocity differences for the redundant pairs, appropriately normalized by the
quadrature sum of the formal velocity errors returned by the program.
We find that the original program errors need to be multiplied by a factor 
of 1.3 in order to match the empirical velocity differences, but 
once that is done a K-S test shows that the resulting normalized 
velocity difference distribution is 
indeed consistent with a Gaussian.
In Figure 8 we plot the velocity difference distribution for the 303 redundant
pairs in the 0223+00 patch;
the rms velocity difference, divided by $\sqrt{2}$, is 103~km~s$^{-1}$,
which indicates our typical random velocity error on a single redshift 
measurement.
We estimate the systematic error in the velocity zero point for our
templates to be approximately 30~km~s$^{-1}$, as detailed in YEC.

\begin{figure}[ht] 
\figurenum{8}
\plotone{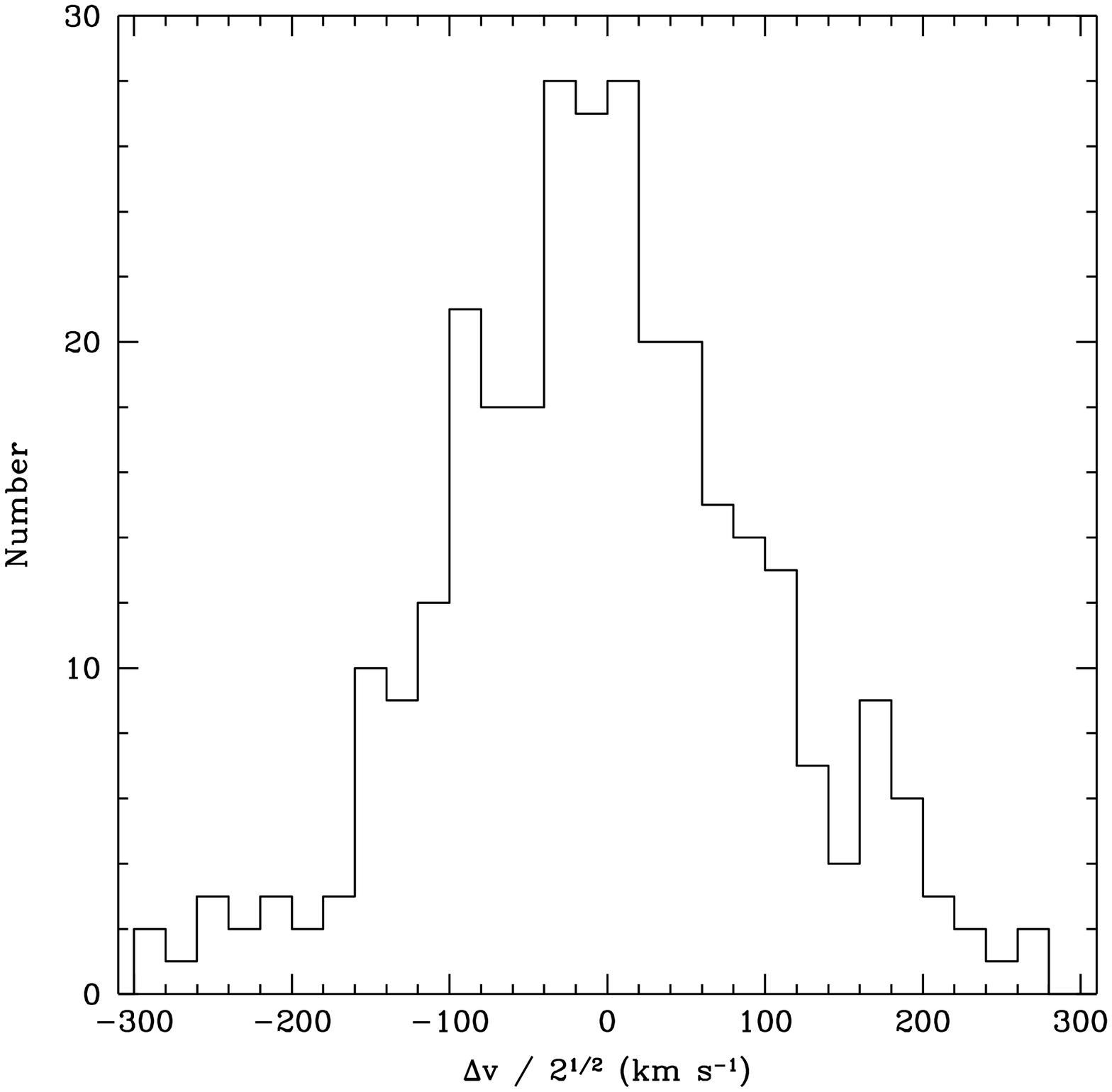} 
\caption{\footnotesize
The distribution of velocity differences for 
redundant $z$ measurements from the CNOC 0223+00 patch scaled
by $\sqrt 2$.
The distribution has a dispersion of 103 km s$^{-1}$.
}

\end{figure}

\section{Completeness and Weights}

The practical difficulty of obtaining spectroscopic observations
for every single galaxy in a faint galaxy sample requires such
a survey to adopt a sparse sampling strategy.
In the case of the CNOC2 survey, the sampling is 
slightly under 2  to 1; and
the sampling itself is not necessarily uniform.
For example, objects near the edges in the spectral direction,
and objects in low galaxy density regions are more likely to be
sampled simply due to geometric considerations.
Furthermore, the success rate of obtaining a redshift when a
galaxy spectrum is obtained depends on many factors, the foremost
being the signal-to-noise ratio and the strength
of the spectral features.
For some galaxies, because of the relatively short spectrum, there
may not be any useful identification features for deriving a robust
redshift.
Hence, to convert an observed redshift sample to a complete sample
requires a detailed understanding of the various effects and methods
of accounting for them.
For the CNOC1 Cluster Redshift Survey, YEC described in detail
their method of compensating
for these various factors by attaching a series of statistical weights
to each galaxy.
The CNOC2 survey uses the same method to assign weights to each
galaxy in the redshift sample.
Here, we will describe very briefly the determination of weights,
with emphasis on the small variations from the method used by YEC.

For each galaxy, a selection function $S=S_mS_{xy}S_cS_z$ is computed,
where $S_m(m)$ is the magnitude ($m$) selection function, $S_{xy}(x,y,m)$
is the geometric selection function based on the location ($x,y$) of the 
object, $S_c(c,m)$ is the color ($c$) selection
function, and $S_z(z,m)$ is the redshift selection function.
The weight, $W$, for each galaxy is then $1/S$.
$S_m$ is chosen as the primary selection function which has a value
between 0 and 1; whereas the other selection functions are considered
as modifiers to $S_m$ and are normalized to have a mean 
over the sample of $\sim 1.0$ (with the exception of $S_z$).
This description of the weights of the galaxies allows one to omit
any of the secondary selection functions if they are deemed 
unnecessary for the analysis.

Selection functions are computed for each filter for every object
with a redshift.
$S_m$ is derived by counting galaxies in a running bin 
of $\pm$0.25 magnitudes around each object, taking the ratio of the number
with redshifts to that in the photometric sample.
Examples of the magnitude selection function for individual
fields are shown in Figure 9 for the patch CNOC$\,$0223+00.
Because there are significant variations of coverage from field
to field due to observing conditions and object density -- the
nominal 20\% selection limit (i.e., $S_m=0.2$) varies from $R_c=21.1$
to 21.8 mag -- the magnitude selection function is computed in
individual fields rather than over the whole patch (which was done
in the CNOC1 catalogs, which have many fewer fields per cluster).
This is equivalent to applying a large geometric filter to the
magnitude selection function.
Variations over smaller angular scale are corrected by $S_{xy}$, which
is computed by counting galaxies within $\pm$0.5 mag
 over an area of 2.0 arcminutes around the object.
The color selection function is similarly accomplished by counting
galaxies over a color range of $\pm$0.25 mag.
The colors used for $I$, $R$, $V$, $B$, and $U$ are $R-I$, $B-R$,
$V-R$, $B-R$, and $U-B$, respectively. 

\begin{figure}[ht] 
\figurenum{9}
\plotone{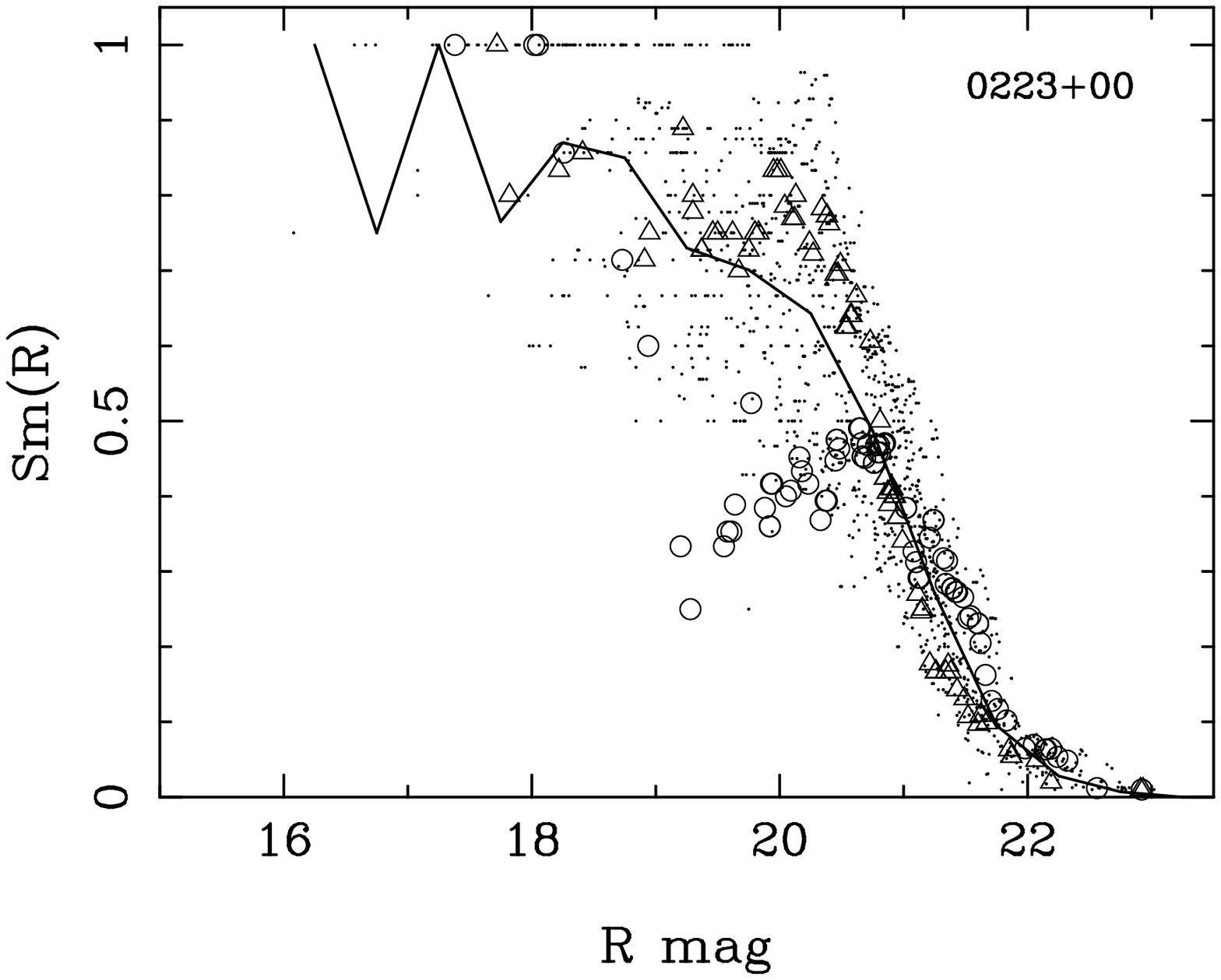} 
\caption{\footnotesize
Magnitude selection function for 
 the fields in the CNOC 0223+00 patch.
Each point represents the magnitude selection value, $S_m$, for the
$R$ band of an object.
The thick solid line represents the mean magnitude selection function 
for the whole patch computed in 1/2 mag bins.
The fluctuations at the bright magnitudes are due to small number
statistics.
To illustrate the variations amongst the fields, the points from
the galaxies in fields {\bf a2} and {\bf c4} are plotted as open circles
and triangles, respectively.
The field {\bf a2} has a particularly low selection for $R<20.5$ mag
galaxies due in part to the relatively poor observing conditions 
for mask A; while the field {\bf c4} is more representative.
}
\end{figure}

The calculations of $S_m$, $S_{xy}$, and $S_c$ are all performed
strictly empirically using the galaxy catalog, and they can
be easily recomputed with different bin sizes, sampling areas,
 or reference colors.
However, the computation of $S_z$ is model dependent, as it requires
an estimate of the number of galaxies of a certain magnitude and color
which are outside the measurable redshift range.
This requires knowledge of the luminosity function of galaxies
as a function of color and redshift, which is in fact one of
the goals in the study of galaxy evolution.
If the LF of the galaxies is known, then a simple correction to
$S_m$, the magnitude selection function, is $f_{z}(z1,z2)/f_{LF}(z1,z2)$;
where $f_{z}(z1,z2)$ is the fraction of galaxies in the redshift
sample between $z1$ and $z2$, and $f_{LF}(z1,z2)$ is the fraction
of galaxies in the same $z$ range as modeled by the LF.
Using the LF derived by Lin et al.~(1999) from two of the 4 CNOC patches,
we can estimate the redshift selection function, $S_z$, as a function
of magnitude.
We note that Lin et al.~derived the LF using a different redshift
weight correction based on colors of galaxies,
although the LF and $S_z$ can be derived iteratively using the above
method for determining $S_z$.
As an example, Table 4 lists $W_z$ derived based on count
models in half magnitude bins for the patch CNOC$\,$0223+00.
The correction to $W_m$ due to the redshift range effect is significant
at $R\gtapr20.5$, rising to as much as 30\% at the nominal spectroscopic
sample limit of $R=21.5$.
A large correction is expected at the faint end, as fainter
galaxies have a higher mean redshift, resulting in fewer galaxies
having their redshift measured.
Values of $W_z$ for individual objects in the catalogs
 are derived by interpolating the results from binned data.
Because of the near 100\% success rate
and small number statistics  at bright magnitudes, $W_z$ is
fixed at 1.0 at $R<18.0$.

Another independent method for accurately accounting for the 
redshift selection 
effect is to compare the redshift sample to a photometric redshift
sample derived from the 5 color photometric data.
This method will be discussed in a future paper.

\begin{figure}[ht] 
\figurenum{11}
\plotone{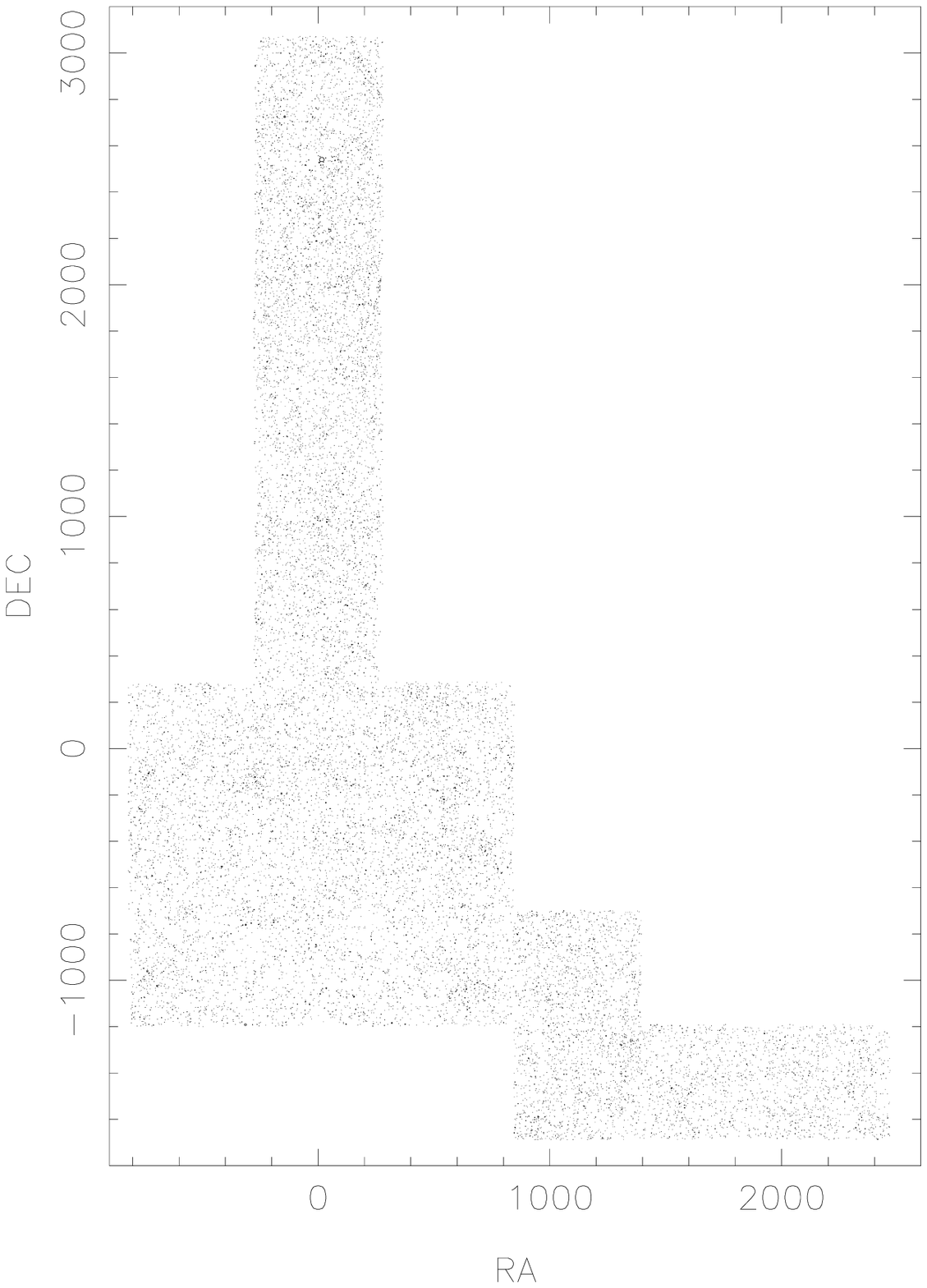} 
\caption{\footnotesize
The photometric sample of the CNOC 0223+00 patch.
All galaxies brighter than 23.0 mag
in $R$ (which is 100\% complete) are plotted.
}
\end{figure}

\section{The Data Catalog for the Patch CNOC$\,$0223+00 } 

The patch CNOC$\,$0223+00 contains 19 fields.
An example of a $R$ MOS image of a single field
 is shown as a gray scale plot in Figure 10, with
objects having measured redshifts marked.
The layout of the 19 fields of the patch is presented in Figure 11, 
with all galaxies
brighter than $R=23.0$ (100\% complete) marked.
The total area in the defined region for the patch is 1408.7 square
arcminutes.
The patch contains to this limit 9554 objects classified as
galaxies and 2574 classified as stars.
The sky and redshift distributions of the
redshift sample are shown in Figures 12 and 13.
A total of 1541 redshifts are measured from 3820 spectrum (with
a redundancy rate of 32.9\%).
The number of galaxies with $R\le21.5$ is 2692, of which
1293 have a measured redshift.
The cumulative sampling rate at $R=21.5$ is 48.0\%, with a raw
success rate of 67.3\%.
Figure 9 illustrates the differential sampling
rate (i.e. $S_m$) of individual objects as a function of magnitude.
The raw and redshift-selection-corrected
success rates as a function of $R$ magnitude are tabulated in Table 4
and plotted in Figure 14.
The corrected success rate is close to 100\% for galaxies
brighter than 20.0, and close to 70\% at the nominal primary spectroscopy
limit of $R=21.5$.
We note that the success rates tabulated reflect the sum of
the A and B masks, with the A masks having half the exposure time
of that of the B masks.
Typical spectra and their correlation functions are plotted in Figure 7.
Figure 8 illustrates the histogram of 
$\Delta v$ from redundant spectroscopic observations. 
The distribution produces an estimate of the
average uncertainty for the redshift measurements of 103 km s$^{-1}$.

\begin{figure}[ht] 
\figurenum{12}
\plotone{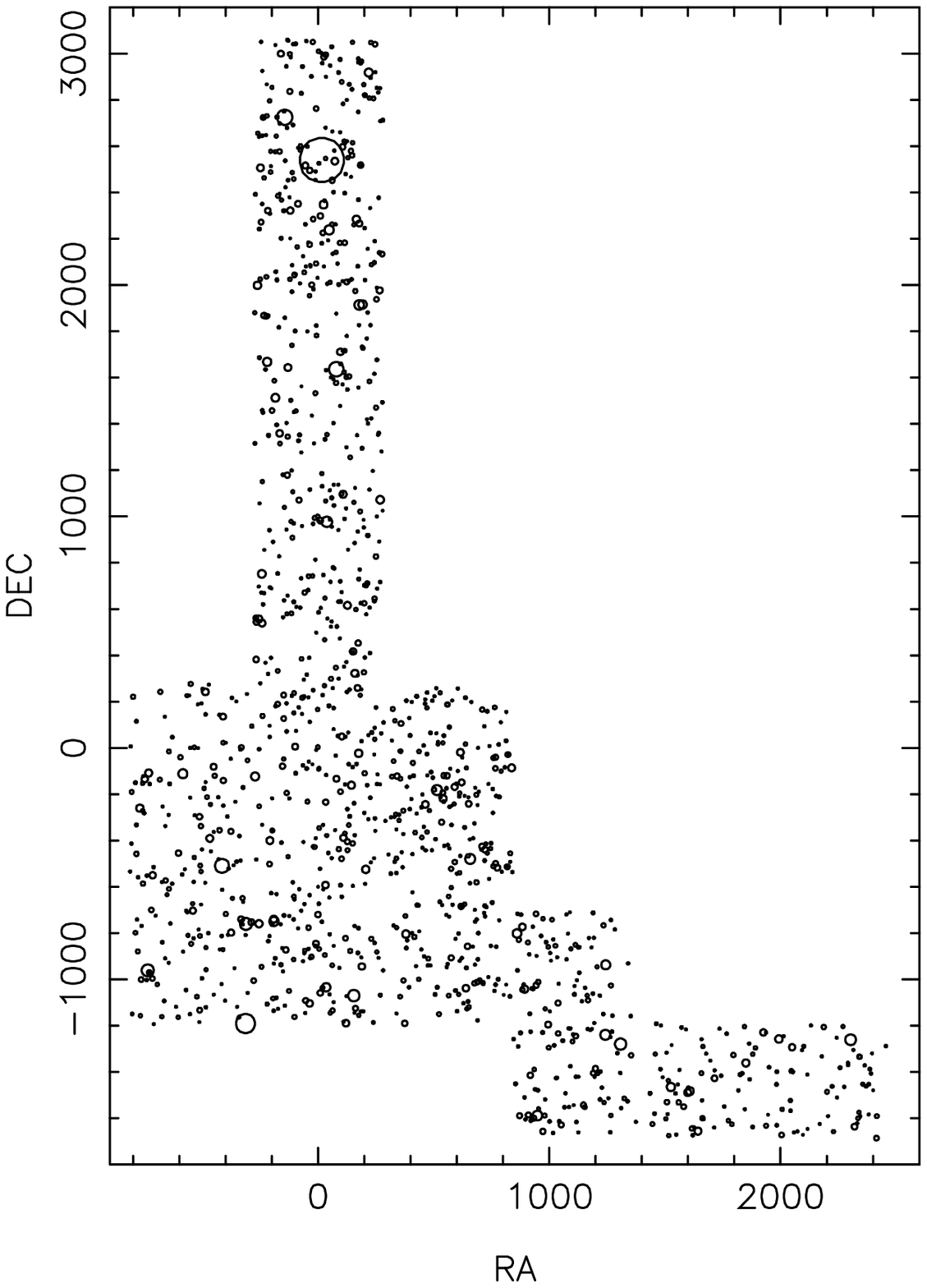} 
\caption{\footnotesize
Same as Figure 12, but for the redshift sample with
$R<21.5$ mag.
The size of the circles is proportional to the apparent $R$ magnitude
of the galaxies.
}
\end{figure}

Some statistics for each field are presented in Table 5.
These include the $5\sigma$ detection limits for 
the $R$ and $B$ filters.
These are fiducial limits determined using stellar PSFs; typical
100\% completeness for galaxies are 0.6 to 0.8 mag brighter
(see Yee 1991).
The numbers in brackets indicate the runs (and hence the CCD used, see
Table 2) from which the $R$ and $B$ images were obtained.
The mean $5\sigma$ magnitude limits are $23.87\pm0.13$ and $24.49\pm0.13$ for
$R$ and $B$, respectively, where the uncertainty is the rms
width of the distribution.
The 100\% photometric completeness limit is approximately 0.7 to 0.9 
magnitudes brighter than the $5\sigma$ limit (see Yee 1991).
For the other 3 filters: $I$, $V$, and $U$, the mean 5$\sigma$ magnitude
limits are 22.80$\pm$0.22, 24.09$\pm$0.12, and 23.04$\pm$0.15, respectively.
Also listed in Table 5 are the number of redshifts measured in each field,
with the bracketed numbers indicating the runs from which masks A, B, and
C were obtained.
The integrated selection function $S_m$ at $R=21.5$ and the
differential selection rate at $21.0<R<21.5$ are also shown in Table 5.
These numbers provide an idea of the variations from field to field
expected in a patch.

\begin{figure}[ht] 
\figurenum{14}
\plotone{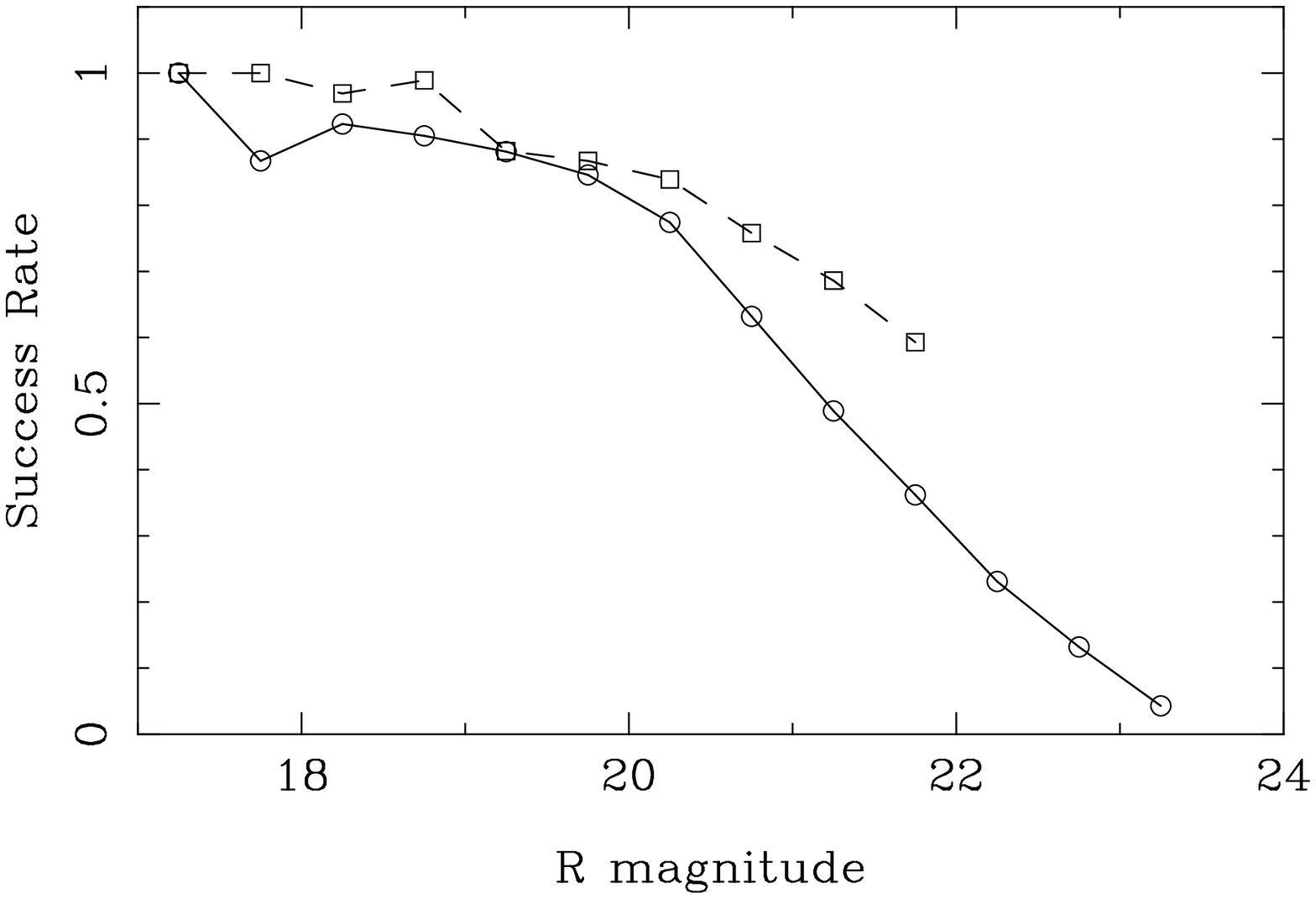} 
\caption{\footnotesize
The raw success rate (solid line) and the redshift-selection
corrected success rate (dashed line) for the CNOC 0223+00 patch.
}
\end{figure}

The fields are merged into a patch by matching a reference object
which appears in both adjacent fields.
Table 6 is a sample catalog for the patch CNOC 0223+00
which  contains every 25th galaxy with redshifts from the full catalog
to demonstrate some of the information tabulated 
for each object in the full catalog.
Listed in the sample catalog are the following columns:

\noindent
Column 1: the PPP object number -- the first two digits represent
the field number, while the last 4 digits is the sequential object
number within the field ordered from South to North. 

\noindent
Column 2: the offset in RA in
arc seconds from the fiducial origin in the {\bf a0} field 
with West being positive. 

\noindent
Column 3: the offset in Dec in arc seconds with North being positive. 
The object positions, both $\Delta$RA and $\Delta$Dec, have not 
been put on a proper astrometric grid.
The relative position between two objects are in general
accurate to 1 to 2 $''$ within the same field, and could
be uncertain by as much as 5$''$ or more across the patch (see
Section 3.1).

\noindent
Columns 4 to 8 : the $I$, $R$, $V$, $B$, and $U$ magnitudes.
The photometric uncertainties for $R$ and $B$ are shown to
illustrate the typical errors.
The errors are tabulated in 1/100 magnitude units.
Note that a 0.03 mag ``aperture'' uncertainty is added in
quadrature to the photon noise error for all objects.
This accounts for the uncertainty in determining the
optimal aperture for the object (See YEC).

\noindent
Column 9: the redshift and uncertainty.
The uncertainty has been scaled by the empirically determined
1.3 factor from that produced by the cross-correlation algorithm
(see Section 4.3) and is tabulated in units of 0.00001 in redshift.

\noindent
Column 10: the spectroscopic classification, with 2=elliptical spectrum,
4=intermediate-type spectrum, 5=emission-line spectrum, 6=active galactic
nuclei.

\noindent
Column 11: cross-correlation coefficient value ($R_{cor}$).

\noindent
Column 12: Spectral energy distribution (SED) class determined
by fitting the 5 color photometry to the empirical SEDs of
galaxies of different spectral types from Coleman, Wu, \& Weedman 
(1980).
The template SEDs of E/S0, Sbc, Scd, and Im classes from
Coleman, Wu, \& Weedman are 
designated as 0.0, 1.0, 2.0, and 3.0.
An additional very blue template, denoted as 4.0, 
is created from the GISSEL library (Bruzual \& Charlot 1996)
for modeling late-type strongly star-forming galaxies.
Intermediate SED classes are interpolations of these templates.

\noindent
Column 13: k-correction for the filter $R$ determined using the
SED class fit and spectral model.

\noindent
Column 14: magnitude weight for the filter $R$.

In the full electronic version of the catalog, the color aperture
error, the k-correction and the
magnitude, geometric, and color weights for the remaining 4 filters,
and the redshift weight for each object are also listed.
The central position in RA and Dec of the catalog 
(i.e., $\Delta$RA=0.0 and $\Delta$Dec=0.0) 
is 00:23:29.2, +00:05:14 (1950).
The complete catalog and detailed explanatory notes for the catalog 
can be found in a number of websites:
http://adc.gsfc.nasa.gov/,
http://www.astro.utoronto.ca/$\sim$hyee/CNOC2/\footnote {
The full electronic version of the catalog will be initially
available at this website on or around August 1, 2000.
}, and 
http://cadcwww.hia.nrc.ca.

\noindent
Also available electronically is a ''field-area map'', which is a
two-dimensional array containing 0's and 1's, mapping the sampled
area of the patch with a 2$''$/pixel resolution.
This map allows the determination of the exact area on the sky
covered, including the effect of blockage by bright stars.


\acknowledgements
We wish to thank the Canadian Time Assignment Committee and the CFHT
for generous allocation of observing time and the CFHT organization for
the technical support.
We especially extend our thanks to the telescope operators: Ken Barton,
John Hamilton, Norman Purves, Dave Woodworth, and Marie-Claire Hainaut 
who, through their dedication and skill, helped
immensely in maximizing the observing efficiency for this large project.
This project was supported by a Collaborative Program grant from
NSERC, as well as individual operating grants from NSERC 
to RGC and HKCY.
HL acknowledges support provided by NASA through Hubble Fellowship grant
\#HF-01110.01-98A awarded by the Space Telescope Science Institute, which 
is operated by the Association of Universities for Research in Astronomy, 
Inc., for NASA under contract NAS 5-26555.

\bigskip
\bigskip

\hoffset=-.20 true in 
\textwidth=7 true in 

\begin{deluxetable}{lccccccc} 
\tablenum{1} 
\tablewidth{7.3 true in} 
\tablecaption{CNOC2 Redshift Survey Patches} 
\tablehead{ 
\colhead{Name} & 
\colhead{RA\tablenotemark{[a]}} & 
\colhead{Dec\tablenotemark{[a]}} & 
\colhead{$b$}  &
\colhead{$E(B-V)$}  &
\colhead{No. of Fields}  &
\colhead{Total Area\tablenotemark{[b]}}  &
\colhead{No. of Masks}  
} 
\tablenotetext{[a]}{Position of the {\bf a0} field, Epoch 1950.}
\tablenotetext{[b]}{Area in square arc minutes.}
\startdata 
CNOC 0223+00   & 02 23 30.0 & +00 06 06 & --54.3 & .036 & 19 & 1408.7 & 39 \\
CNOC 0920+37   & 09 20 40.7 & +37 18 17 & +45.6 & .012 & 19 & 1366.9 & 40 \\
CNOC 1447+09   & 14 47 11.6 & +09 21 21 & +57.2 & .029 & 17 & 1240.0 & 34 \\
CNOC 2148--05   & 21 48 43.0 & --05 47 32 & --41.6 & .035 & 19 & 1417.9 & 39 \\

\enddata
\end{deluxetable}


\begin{deluxetable}{lccc} 
\tablenum{2} 
\tablewidth{6 true in} 
\tablecaption{Properties of CCD Detectors and Exposure Times} 
\tablehead{ 
\colhead{} & 
\colhead{ORBIT1} & 
\colhead{LORAL3} & 
\colhead{STIS2}  
} 

\startdata 
Pixel size ($\mu$m)      & 15  & 15  & 21 \\
Pixel scale ($''$)       & 0.313  & 0.313  & 0.438 \\
Imaging Area Used (pixels)   & $2048\times 2048$ & $2048\times 2048$ & $1416\times 1368$    \\
Defined Field Area ($'$) & $7.3 \times 9.2$ & $7.3 \times 9.2$ 
                   & $8.3 \times 9.2$     \\
\AA~per pixel            &  3.55     &  3.55   & 4.96   \\
Spectroscopy Exp. Time (sec)   &         &        &      \\
~~~~Mask A   & $\sim$3000   &  $\sim$3600       & 2400 \\
~~~~Mask B   & $\sim$7000    &   $\sim$7200      & 4800 \\
Imaging Exp. Time (sec) &         &        &      \\
~~~~Filter $I_c$   & 420     &  600   & 360  \\
~~~~Filter $R_c$   & 420     &  600   & 420  \\
~~~~Filter $V$   &   420   &    600 & 420  \\
~~~~Filter $B$   &   420   &    900 & 480  \\
~~~~Filter $U$   &  not used  &  not used   & 900  \\

\enddata
\end{deluxetable}

\begin{deluxetable}{lccl} 
\tablenum{3} 
\tablewidth{5 true in} 
\tablecaption{Journal of Observations}
\tablehead{ 
\colhead{Run Number} & 
\colhead{Beginning Date (UT)} & 
\colhead{Number of Nights} & 
\colhead{CCD}  
} 

\startdata 
1.     & 1995/02/27  & 8  & ORBIT1 \\
2.     & 1995/10/20  & 8  & LORAL3 \\
3.     & 1996/02/13  & 9  & LORAL3 \\
4.     & 1996/10/11  & 8  & STIS2  \\
5.     & 1997/02/04  & 8  & STIS2  \\
6.     & 1997/08/30  & 6  & STIS2  \\
7.     & 1998/05/23  & 6  & STIS2  \\

\enddata
\end{deluxetable}

\begin{deluxetable}{cccccc} 
\tablenum{4} 
\tablewidth{6.5 true in} 
\tablecaption{Redshift Selection Function and Success Rates
for Patch CNOC 0223+00}
\tablehead{ 
\colhead{Mag bin} & 
\colhead{$f_z(0.12,0.55)$} & 
\colhead{$f_{LF}(0.12,0.55)$ } & 
\colhead{$W_z$} & 
\colhead{Raw Success Rate } &
\colhead{Corrected Success Rate}
} 
\startdata 
17.0--17.5 & 0.667 & 0.500  & 0.752  & 1.00 & 1.33 \\
17.5--18.0 & 0.923 & 0.687  & 0.746  & 0.87 & 1.16 \\
18.0--18.5 & 0.854 & 0.814  & 0.953  & 0.92 & 0.97 \\
18.5--19.0 & 0.970 & 0.887  & 0.915  & 0.91 & 0.99 \\
19.0--19.5 & 0.928 & 0.927  & 0.999  & 0.88 & 0.88 \\
19.5--20.0 & 0.950 & 0.927  & 0.976  & 0.85 & 0.87 \\
20.0--20.5 & 0.942 & 0.869  & 0.922  & 0.77 & 0.84 \\
20.5--21.0 & 0.903 & 0.752  & 0.833  & 0.63 & 0.76 \\
21.0--21.5 & 0.853 & 0.608  & 0.713  & 0.49 & 0.69 \\
21.5--22.0 & 0.774 & 0.472  & 0.610  & 0.36 & 0.59 \\
\enddata
\end{deluxetable}

\begin{deluxetable}{ccccccc} 
\tablenum{5} 
\tablewidth{6.5 true in} 
\tablecaption{Statistics for Patch CNOC 0223+00}
\tablehead{ 
\colhead{Field Number} & 
\colhead{Field Name} & 
\colhead{$R_c$\tablenotemark{[a]}} & 
\colhead{$B$\tablenotemark{[a]}} & 
\colhead{$N_z$} & 
\colhead{$S_{R<21.5}$}  &
\colhead{$S_{21.0<R<21.5}$}  
} 
\tablenotetext{[a]}{$5\sigma$ detection magnitude, the number in the
bracket denotes the run number.}
\startdata 
1 & a0  & 24.02 (2)  & 24.55 (4)  & 55 (2,2,4) & 0.399 & 0.260 \\
2 & a1  & 24.02 (2)  & 24.35 (4)  & 62 (2,2)~~ & 0.512 & 0.327 \\
3 & a2  & 24.04 (2)  & 24.56 (4)  & 70 (2,4)~~ & 0.424 & 0.400 \\
4 & a3  & 23.95 (4)  & 24.60 (4)  & 57 (2,4)~~ & 0.518 & 0.358 \\
5 & a4  & 23.86 (4)  & 24.63 (4)  & 70 (4,4)~~ & 0.387 & 0.164 \\
6 & a5  & 24.02 (4)  & 24.55 (4)  & 78 (4,4)~~ & 0.382 & 0.151 \\
7 & a6  & 23.93 (6)  & 24.57 (6)  & 88 (6,6)~~ & 0.550 & 0.236 \\
8 & b1  & 23.78 (4)  & 24.53 (4) &  49 (4,4)~~ & 0.383 & 0.255 \\
9 & b2  & 23.64 (4)  & 24.38 (4)  & 65 (4,4)~~ & 0.489 & 0.373 \\
10 & b3  & 23.83 (4)  & 24.17 (4) & 60 (4,4)~~ & 0.472 & 0.315 \\
11 & b4  & 23.67 (4)  & 24.51 (4)  & 77 (4,4)~~ & 0.653 & 0.475 \\
12 & b5  & 23.78 (4)  & 24.47 (4)  &  62 (4,4)~~ & 0.508 & 0.255 \\
13 & c1  & 23.87 (4)  & 24.59 (4)  & 87 (4,4)~~  & 0.503 & 0.279 \\
14 & c2  & 23.76 (4)  & 24.32 (4)  & 97 (4,4)~~  & 0.508 & 0.293 \\
15 & c3  & 23.63 (4)  & 24.47  (4) & 60 (4,4)~~  & 0.492 & 0.213 \\
16 & c4  & 23.93 (4)  & 24.60 (4)  & 64 (5,5)~~  & 0.454 & 0.172 \\
17 & c5  & 23.92 (4)  & 24.66 (4)  & 72 (5,5)~~ &  0.500 & 0.151 \\
18 & c6  & 23.90 (4)  & 24.27 (4)  & 61 (5,6)~~ &  0.535 & 0.216 \\
19 & c7  & 24.04 (6)  & 24.46 (6)  & 59 (6,6)~~ &  0.590 & 0.378 \\
   &     &   &  &  &  & \\
Total & & & & 1293~~~~ & 0.480 & 0.270 \\
\enddata
\end{deluxetable} 

\newpage


\begin{deluxetable}{lrrcccccccrccc} 
\tablenum{6} 
\tablewidth{8.4true in} 
\tablecaption{Sample Catalog for the CNOC 0223+00 Patch}
\rotate
\tablehead{ 
\colhead{PPP\#} & 
\colhead{$\Delta$ RA} & 
\colhead{$\Delta$ Dec} & 
\colhead{$I_c$ } & 
\colhead{$R_c$ } & 
\colhead{$V$} & 
\colhead{$B$} & 
\colhead{$U$} & 
\colhead{$z$} & 
\colhead{Scl} & 
\colhead{$R_{cor}$} & 
\colhead{SED} & 
\colhead{K($R_c$)} & 
\colhead{$W_m$}  
} 

\startdata

062246 &    16.10 &  2540.30 & 13.62 & 14.25$\pm$03 &14.89 & 15.89$\pm$03 &16.22 & 
           0.02383$\pm$034 & 5 &  4.56 & 0.20 &  0.02 & 1.00 \\
071667 &   219.00 &  2918.60 & 17.21 & 17.97$\pm$03 &18.86 & 20.59$\pm$04 & 0.00 & 
           0.25109$\pm$029 & 2 & 11.10 & 0.08 &  0.29 & 1.25 \\
120965 &  -208.60 &  -400.20 & 17.77 & 18.28$\pm$03 &18.71 & 19.37$\pm$03 &19.43 & 
           0.09095$\pm$036 & 4 &  3.98 & 2.02 &  0.01 & 1.00 \\
182286 &  1925.80 & -1229.50 & 17.85 & 18.50$\pm$03 &18.94 & 19.90$\pm$03 &19.63 & 
           0.14918$\pm$036 & 5 &  4.03 & 1.49 &  0.05 & 1.00 \\
031395 &     9.50 &   982.70 & 18.03 & 18.73$\pm$03 &19.49 & 20.72$\pm$04 &20.54 & 
           0.26713$\pm$031 & 4 &  6.32 & 0.66 &  0.21 & 1.40 \\
061186 &  -217.50 &  2321.70 & 18.20 & 18.87$\pm$03 &19.47 & 20.58$\pm$04 &20.19 & 
           0.23756$\pm$034 & 4 &  5.49 & 1.27 &  0.10 & 1.50 \\
170596 &   871.70 & -1589.60 & 18.24 & 19.05$\pm$03 &20.05 & 21.56$\pm$04 &21.79 & 
           0.28857$\pm$039 & 2 &  7.98 & 0.07 &  0.34 & 1.09 \\
120706 &   101.50 &  -478.90 & 18.44 & 19.17$\pm$03 &19.97 & 21.13$\pm$04 &21.00 & 
           0.35873$\pm$042 & 4 &  3.86 & 0.88 &  0.25 & 1.17 \\
172593 &  1038.20 & -1233.90 & 18.48 & 19.29$\pm$03 &19.96 & 21.31$\pm$04 &21.41 & 
           0.27141$\pm$030 & 5 &  8.61 & 0.63 &  0.23 & 1.50 \\
020465 &   198.00 &   326.80 & 18.78 & 19.38$\pm$03 &19.83 & 20.83$\pm$04 &20.43 & 
           0.25072$\pm$029 & 5 &  9.52 & 2.14 &  0.00 & 1.80 \\
130656 &   758.30 &   -43.30 & 18.67 & 19.49$\pm$03 &20.57 & 22.28$\pm$08 &22.91 & 
           0.30653$\pm$033 & 2 &  6.04 &-0.04 &  0.39 & 1.40 \\
110674 &  -244.50 &  -945.70 & 18.57 & 19.60$\pm$03 &20.80 & 22.43$\pm$08 &23.40 & 
           0.50610$\pm$130 & 2 &  7.00 & 0.27 &  0.77 & 1.17 \\
101269 &  -511.00 &  -812.60 & 18.94 & 19.68$\pm$03 &20.56 & 21.94$\pm$06 &22.30 & 
           0.23808$\pm$029 & 2 &  9.35 & 0.21 &  0.25 & 1.12 \\
072259 &   228.20 &  3048.20 & 18.93 & 19.75$\pm$04 &20.69 & 22.03$\pm$06 & 0.00 & 
           0.38166$\pm$055 & 4 &  3.70 & 0.57 &  0.38 & 1.47 \\
061882 &  -233.80 &  2463.40 & 18.93 & 19.82$\pm$04 &20.92 & 22.49$\pm$08 &23.18 & 
           0.34269$\pm$039 & 2 &  4.31 & 0.06 &  0.45 & 1.57 \\
121045 &   115.80 &  -369.20 & 19.18 & 19.88$\pm$04 &20.61 & 21.78$\pm$05 &21.63 & 
           0.26797$\pm$030 & 4 &  7.42 & 0.79 &  0.19 & 1.71 \\
060425 &   -90.60 &  2164.80 & 19.26 & 19.93$\pm$04 &20.73 & 21.74$\pm$05 &21.38 & 
           0.39500$\pm$052 & 4 &  2.74 & 1.53 &  0.17 & 1.67 \\
140958 &   731.80 &  -454.60 & 19.18 & 19.98$\pm$04 &21.13 & 22.70$\pm$08 &23.38 & 
           0.35861$\pm$035 & 2 &  5.58 & 0.23 &  0.44 & 1.41 \\
131723 &   527.80 &   238.40 & 19.53 & 20.02$\pm$04 &20.42 & 21.24$\pm$04 &21.03 & 
           0.20448$\pm$027 & 5 & 11.25 & 2.52 & -0.04 & 1.35 \\
041010 &   -80.40 &  1353.00 & 19.61 & 20.09$\pm$04 &20.62 & 21.32$\pm$04 &21.16 & 
           0.19236$\pm$034 & 5 &  5.26 & 2.28 & -0.01 & 1.27 \\
141131 &   724.90 &  -414.70 & 19.51 & 20.14$\pm$04 &20.98 & 22.04$\pm$06 &21.60 & 
           0.40983$\pm$031 & 5 &  6.84 & 1.56 &  0.17 & 1.61 \\
191280 &  2355.00 & -1452.90 & 19.66 & 20.19$\pm$04 &20.77 & 21.76$\pm$05 &21.41 & 
           0.34958$\pm$031 & 5 &  7.50 & 2.20 &  0.03 & 1.19 \\
062186 &     3.20 &  2526.10 & 19.66 & 20.23$\pm$05 &21.08 & 22.06$\pm$07 &22.75 & 
           0.47153$\pm$040 & 5 &  4.16 & 0.98 &  0.34 & 2.22 \\
071950 &   174.10 &  2978.30 & 19.43 & 20.29$\pm$04 &21.50 & 22.90$\pm$11 & 0.00 & 
           0.47065$\pm$038 & 4 &  5.26 & 0.48 &  0.58 & 1.16 \\
040918 &    29.70 &  1333.40 & 19.64 & 20.32$\pm$04 &21.27 & 21.82$\pm$06 &21.53 & 
           0.57002$\pm$031 & 5 &  7.97 & 2.16 &  0.23 & 1.50 \\
090572 &  -352.60 &  -491.60 & 19.70 & 20.37$\pm$04 &21.05 & 22.11$\pm$06 &21.83 & 
           0.36132$\pm$030 & 5 &  8.19 & 1.69 &  0.12 & 2.27 \\
141456 &   576.80 &  -332.70 & 19.49 & 20.41$\pm$05 &21.68 & 23.27$\pm$10 &24.40 & 
           0.43243$\pm$036 & 2 &  5.12 & 0.16 &  0.61 & 2.21 \\
172640 &  1270.80 & -1225.90 & 19.64 & 20.45$\pm$04 &21.14 & 22.51$\pm$06 &22.45 & 
           0.30501$\pm$030 & 5 &  7.76 & 0.69 &  0.25 & 1.12 \\
120172 &  -103.10 &  -653.80 & 20.03 & 20.48$\pm$04 &20.88 & 21.69$\pm$05 &21.42 & 
           0.21864$\pm$027 & 5 & 12.80 & 2.62 & -0.05 & 1.73 \\
160180 &  1273.00 & -1157.20 & 19.58 & 20.54$\pm$04 &21.57 & 22.94$\pm$10 &23.26 & 
           0.38611$\pm$017 & 5 &  7.60 & 0.29 &  0.48 & 1.85 \\
081751 &  -499.30 &   254.00 & 19.67 & 20.58$\pm$04 &21.36 & 23.33$\pm$10 &25.75 & 
           0.38505$\pm$033 & 2 &  6.60 & 0.25 &  0.48 & 3.12 \\
170502 &   896.60 & -1605.40 & 19.70 & 20.62$\pm$04 &21.91 & 23.35$\pm$11 &24.21 & 
           0.38629$\pm$035 & 2 &  5.70 & 0.03 &  0.55 & 1.48 \\
172734 &  1090.10 & -1212.20 & 20.14 & 20.66$\pm$04 &21.32 & 22.22$\pm$06 &21.67 & 
           0.44641$\pm$030 & 5 &  8.21 & 2.42 &  0.04 & 1.52 \\
061970 &   118.80 &  2479.40 & 20.29 & 20.70$\pm$04 &21.30 & 22.03$\pm$06 &21.67 & 
           0.40574$\pm$029 & 5 & 10.51 & 2.67 & -0.03 & 2.67 \\
031433 &   114.20 &   989.70 & 20.33 & 20.76$\pm$04 &21.28 & 22.19$\pm$06 &21.83 & 
           0.42381$\pm$031 & 5 &  7.71 & 2.50 &  0.01 & 2.37 \\
140585 &   338.70 &  -541.00 & 19.85 & 20.78$\pm$04 &21.95 & 23.94$\pm$29 &23.94 & 
           0.44481$\pm$039 & 2 &  4.69 & 0.23 &  0.62 & 1.75 \\
131602 &   494.50 &   204.30 & 19.95 & 20.82$\pm$04 &22.03 & 23.76$\pm$13 &24.79 & 
           0.50480$\pm$130 & 1 &  4.00 & 0.42 &  0.69 & 2.69 \\
110893 &   -41.40 &  -868.70 & 20.03 & 20.85$\pm$05 &21.54 & 22.60$\pm$08 &22.49 & 
           0.30240$\pm$013 & 5 &  8.64 & 0.83 &  0.21 & 1.87 \\
140073 &   727.80 &  -688.20 & 20.28 & 20.89$\pm$04 &21.67 & 22.40$\pm$08 &21.80 & 
           0.61474$\pm$046 & 5 &  3.98 & 2.42 &  0.21 & 1.66 \\
150997 &   560.00 &  -933.50 & 20.08 & 20.93$\pm$04 &22.02 & 23.70$\pm$17 &23.79 & 
           0.40895$\pm$056 & 2 &  2.57 & 0.36 &  0.50 & 2.11 \\
150326 &   628.80 & -1102.70 & 20.10 & 20.96$\pm$04 &22.09 & 23.88$\pm$19 &26.84 & 
           0.38629$\pm$036 & 2 &  5.48 & 0.05 &  0.54 & 2.23 \\
131156 &   718.70 &    93.30 & 20.46 & 21.01$\pm$04 &21.77 & 22.64$\pm$07 &22.19 & 
           0.39897$\pm$031 & 5 &  6.94 & 2.12 &  0.07 & 2.44 \\
140971 &   361.50 &  -450.80 & 20.47 & 21.05$\pm$05 &21.69 & 22.67$\pm$09 &21.96 & 
           0.36204$\pm$029 & 5 & 10.29 & 2.20 &  0.04 & 2.29 \\
151247 &   669.30 &  -859.10 & 20.47 & 21.09$\pm$05 &21.62 & 22.43$\pm$07 &22.17 & 
           0.35917$\pm$027 & 5 & 16.01 & 2.38 &  0.01 & 2.30 \\
041846 &   174.70 &  1556.40 & 20.54 & 21.13$\pm$05 &21.73 & 22.16$\pm$06 &21.88 & 
           0.52669$\pm$034 & 5 &  6.50 & 3.00 & -0.07 & 2.48 \\
101399 &  -326.20 &  -762.70 & 20.31 & 21.16$\pm$06 &22.27 & 23.18$\pm$16 &22.28 & 
           0.58764$\pm$034 & 5 &  6.73 & 1.44 &  0.43 & 2.79 \\
040195 &    87.10 &  1137.70 & 20.57 & 21.19$\pm$05 &22.12 & 23.25$\pm$10 &23.09 & 
           0.42623$\pm$034 & 5 &  6.00 & 1.20 &  0.25 & 2.65 \\
091130 &  -646.60 &  -290.40 & 20.77 & 21.24$\pm$08 &21.82 & 22.92$\pm$10 &22.34 & 
           0.35491$\pm$029 & 5 & 10.63 & 2.22 &  0.03 & 2.79 \\
160798 &  1109.30 & -1028.80 & 20.85 & 21.26$\pm$05 &21.75 & 22.34$\pm$07 &22.14 & 
           0.18766$\pm$027 & 5 & 17.35 & 2.74 & -0.06 & 6.00 \\
110372 &    49.10 & -1065.10 & 20.49 & 21.31$\pm$08 &22.20 & 23.25$\pm$09 &23.77 & 
           0.44648$\pm$012 & 5 & 11.53 & 0.83 &  0.38 & 2.33 \\
041540 &   -92.20 &  1471.70 & 20.39 & 21.37$\pm$05 &22.50 & 24.49$\pm$16 &25.92 & 
           0.47990$\pm$048 & 4 &  3.42 & 0.14 &  0.75 & 3.00 \\
081385 &  -609.90 &   138.20 & 20.89 & 21.43$\pm$06 &21.99 & 99.00$\pm$99 &22.70 & 
           0.38410$\pm$030 & 5 &  8.16 & 2.24 &  0.04 & 5.91 \\
120096 &  -167.00 &  -679.70 & 20.98 & 21.49$\pm$07 &22.37 & 23.17$\pm$10 &22.29 & 
           0.47115$\pm$055 & 5 &  3.05 & 2.33 &  0.08 & 6.27 \\
120867 &   -84.60 &  -428.00 & 21.01 & 21.55$\pm$06 &22.11 & 22.80$\pm$08 &22.14 & 
           0.65007$\pm$031 & 5 &  9.45 & 2.98 &  0.05 & 8.75 \\
090545 &  -451.20 &  -504.20 & 21.32 & 21.62$\pm$08 &22.10 & 23.08$\pm$15 &23.81 & 
           0.20796$\pm$031 & 5 &  6.12 & 1.84 &  0.03 & 5.14 \\
130560 &   658.40 &   -68.10 & 20.93 & 21.68$\pm$06 &22.37 & 22.84$\pm$10 &22.23 & 
           0.61448$\pm$029 & 5 & 10.87 & 2.76 &  0.09 &32.33 \\
010558 &  -225.10 &  -102.40 & 21.18 & 21.76$\pm$06 &22.57 & 23.24$\pm$09 &22.52 & 
           0.58774$\pm$031 & 5 &  7.93 & 2.51 &  0.14 &13.50 \\
090794 &  -326.50 &  -414.70 & 21.24 & 21.86$\pm$08 &22.59 & 23.30$\pm$12 &22.76 & 
           0.65381$\pm$029 & 5 & 10.98 & 2.39 &  0.27 &13.83 \\
150071 &   695.60 & -1179.20 & 21.48 & 22.00$\pm$10 &22.42 & 23.09$\pm$10 &22.53 & 
           0.42566$\pm$030 & 5 &  8.98 & 3.21 & -0.13 &17.67 \\
041681 &   132.70 &  1510.90 & 21.62 & 22.15$\pm$10 &22.63 & 23.26$\pm$08 &22.90 & 
           0.50800$\pm$034 & 5 &  6.43 & 3.00 & -0.07 &57.00 \\
011670 &   170.20 &   176.70 & 22.02 & 22.32$\pm$09 &22.70 & 23.73$\pm$15 &23.16 & 
           0.29867$\pm$031 & 5 &  6.39 & 2.80 & -0.07 &25.25 \\
110833 &    59.20 &  -891.10 & 22.34 & 22.68$\pm$11 &23.31 & 24.61$\pm$31 &24.45 & 
           0.35762$\pm$026 & 5 &  4.86 & 1.85 &  0.09 &26.20 \\

\enddata
\end{deluxetable}
\newpage
\hoffset=0truein

\onecolumn
\begin{figure}[ht] \figurenum{7a}
\plotone{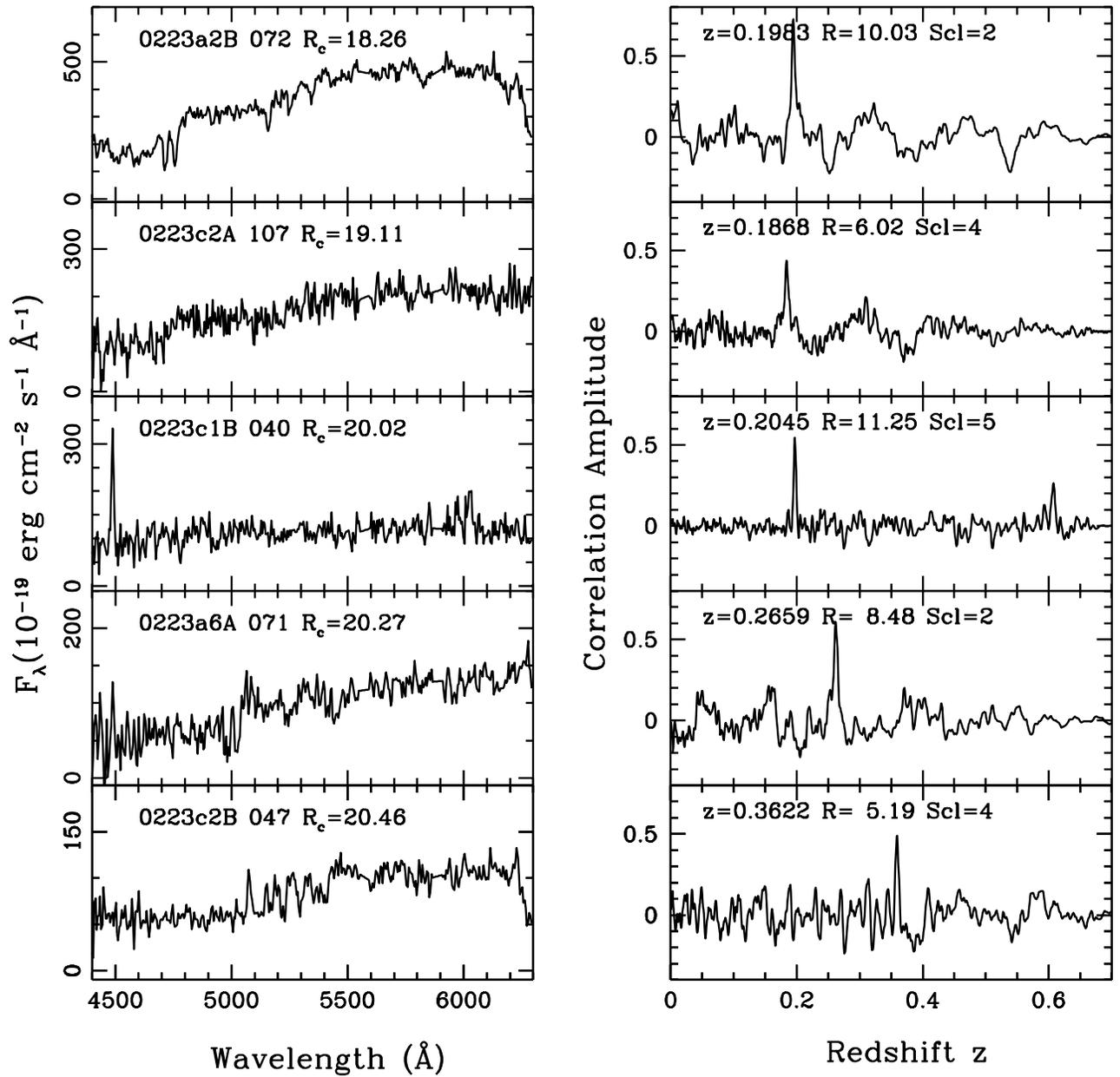} 
\caption{\footnotesize
Examples of galaxy spectra and correlation functions.
Objects are selected randomly from the catalog and plotted in
order of their $R$ magnitudes.
The text on the spectral plot (left panel) indicates the field name,
mask name, slit number and $R$ magnitude, while that
on the correlation plot (right panel) indicates the redshift,
the correlation coefficient $R_{cor}$, and the spectral class
designation SCl (see text).
}
\end{figure}
\newpage
\begin{figure}[ht] 
\figurenum{7b}
\plotone{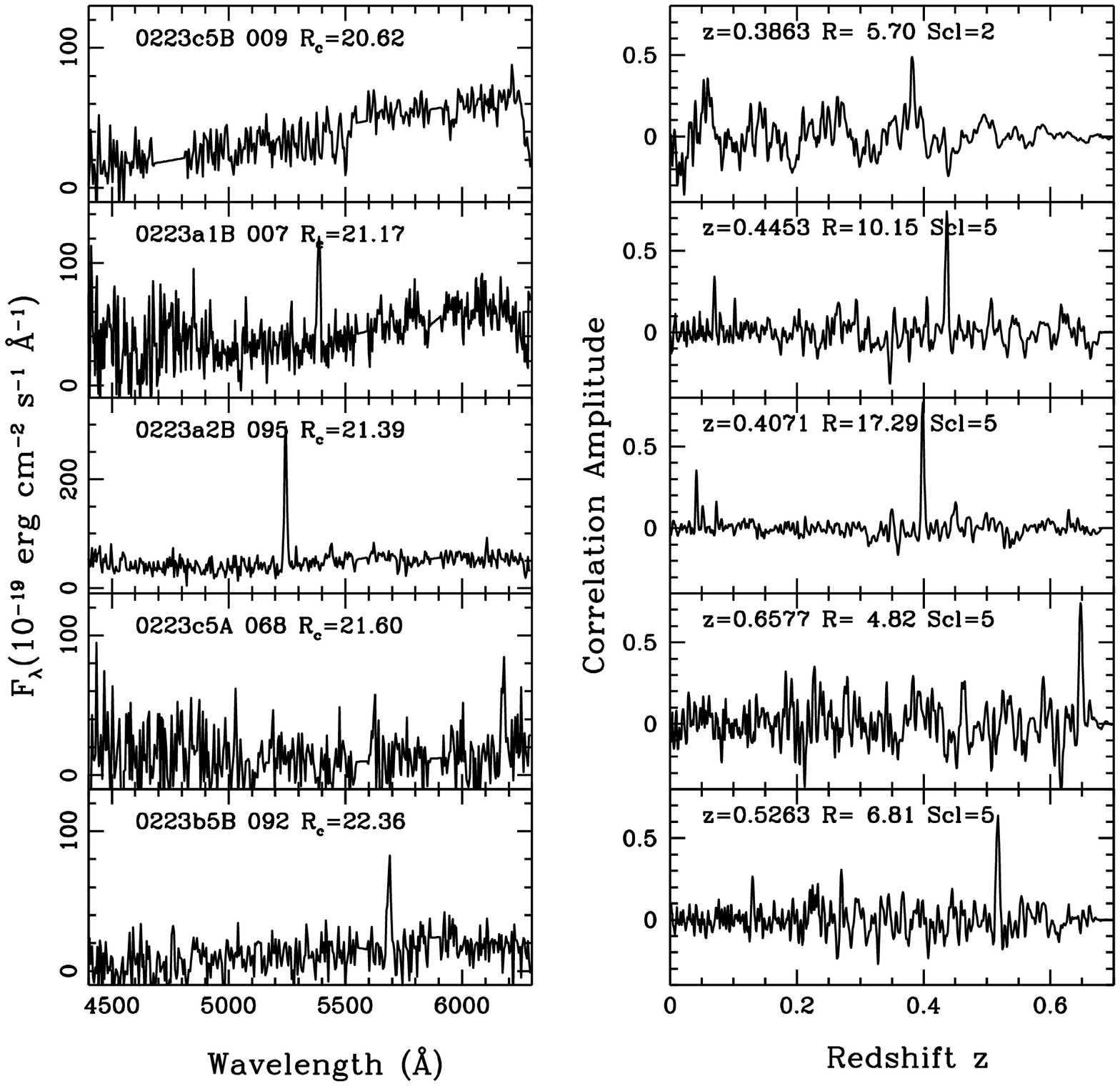} 
\caption{}
\end{figure}

\begin{figure}[ht] 
\figurenum{10}
\plotone{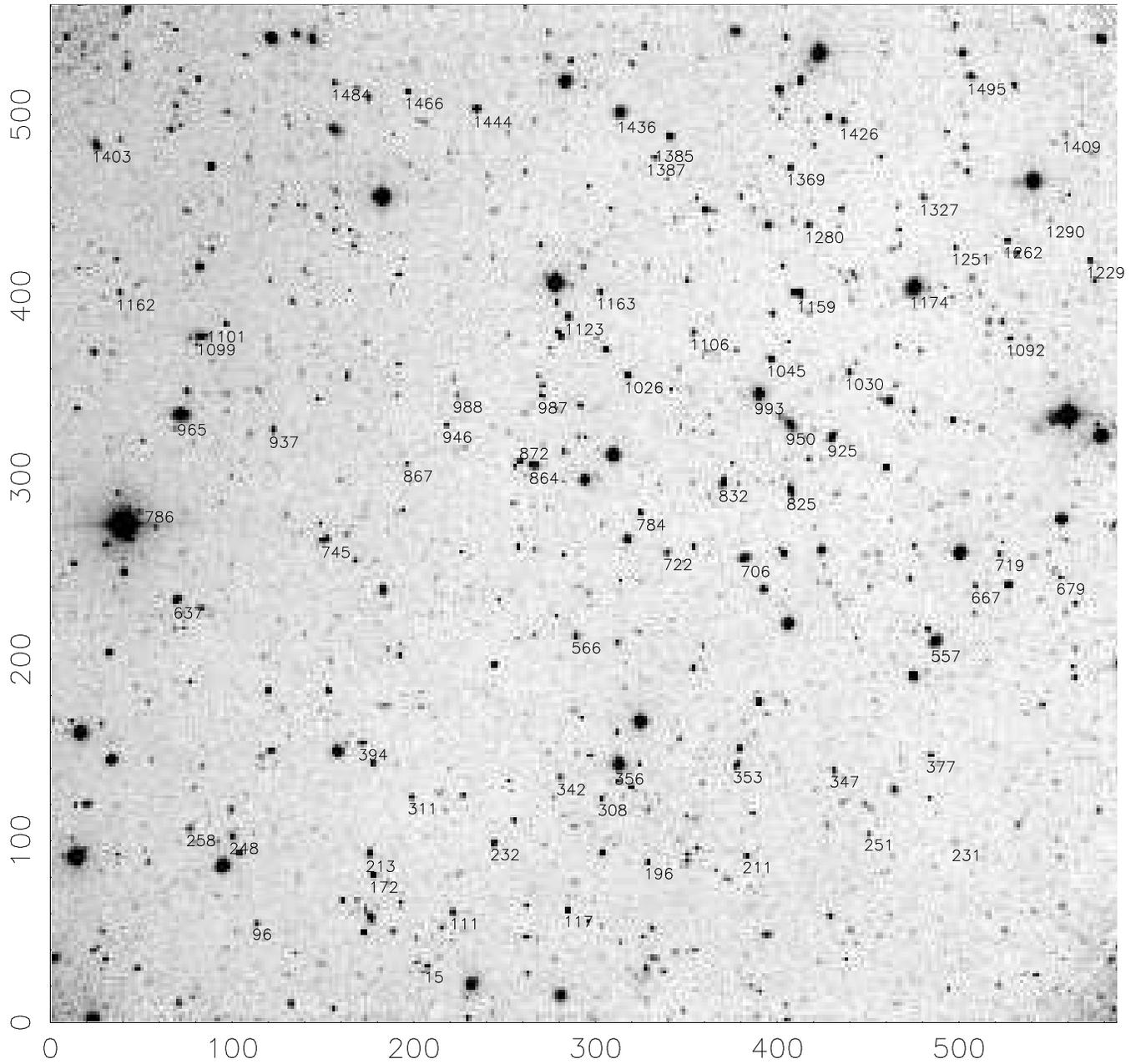} 
\caption{
An example of single field $R$-band images 
from the survey plotted as gray scales.
The field is CNOC 0223+00{\bf b5}.
North-East is upper left, with the axes marked in 100 arcsecond
intervals.
The total field size is $\sim 10'\times 10'$.
The PPP object numbers of objects with measured redshifts are shown.
(Note: This version of the image has been block averaged to reduce the file size.)
}

\end{figure}

\begin{figure}[h] 
\figurenum{13}
\plotone{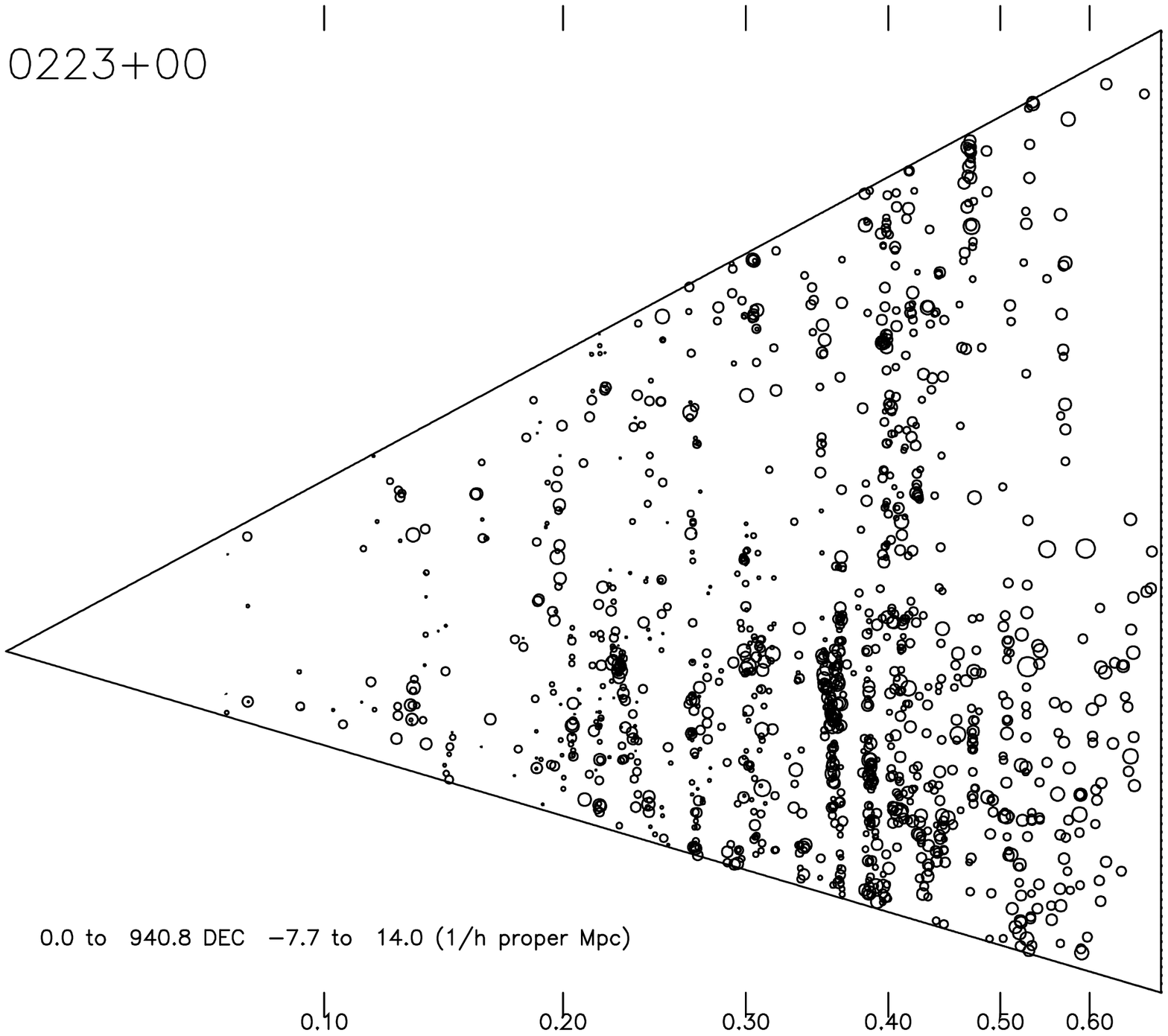} 
\caption{
Cone diagram for the CNOC 0223+00 patch, projected along 
the Declination axis and plotted in units of proper Mpc.
The x-axis  is marked in intervals of 0.1 in redshift,
representing a total length of 940.8 $h^{-1}$ Mpc.
The cone diagram's true opening angle is 1.3$^{\rm o}$, but it
is expanded to make the galaxy points visible.
The y-axis has a total length of 21.7 $h^{-1}$ Mpc at $z=0.7$.
The vertex of the cone is placed at the center of the {\bf a0} field.
The size of the circles indicates the relative luminosity of the
galaxies.
The apparent changes in the density of galaxies along 
the vertical axis is due to a different 
number of fields being projected at different declinations.
}

\end{figure}

\newpage

\end{document}